\documentclass[sigplan,10pt]{acmart}
\renewcommand\footnotetextcopyrightpermission[1]{}

\usepackage[ruled,linesnumbered]{algorithm2e} % For algorithms

\usepackage{amsmath}

\usepackage{graphicx}
\usepackage{paralist}

\usepackage[subtle]{savetrees}
\usepackage[compact]{titlesec}

\titlespacing{\section}{0pt}{1.5ex}{0.5ex}
\titlespacing{\subsection}{0pt}{1ex}{0.5ex}
\titlespacing{\subsubsection}{0pt}{0.5ex}{0.5ex}
\titlespacing{\paragraph}{0pt}{0.5ex}{1.5ex}

\usepackage{hyperref}

\usepackage{bbding}
\usepackage{tabularx}

\newcommand{\greencheck}{ {\color{green}\checkmark} }
\newcommand{\redx}{ {\color{red}\XSolidBold} }
\newcommand{\contracttableyes}{\greencheck}
\newcommand{\contracttableno}{\redx}

\def\tx{\textsf{tx}}

\usepackage{caption}
\ifnum\pdfstrcmp{\jobname}{draft}=0

\newcommand{\TODO}[1]{\begingroup \bfseries\color{red} TODO #1\endgroup}
\newcommand{\anonymize}{0}
\newcommand{\shrink}{0}
\fi

\ifnum\pdfstrcmp{\jobname}{anonymized}=0

\newcommand{\TODO}[1]{}
\newcommand{\anonymize}{1}
\newcommand{\shrink}{1}
\fi

\ifnum\pdfstrcmp{\jobname}{main}=0

\newcommand{\TODO}[1]{}
\newcommand{\anonymize}{0}
\newcommand{\shrink}{0}
\fi

\if\anonymize1
\newcommand{\githuburl}{<redacted>}
\else
\newcommand{\githuburl}{\url{https://github.com/scslab/smart-contract-scalability}}
\fi

\newcommand{\codefontsize}{\small}

\if\shrink1

\newenvironment{geoffitemize}
{
	\begin{compactitem}
}
{ 
	\end{compactitem}
}

\fi
\if\shrink0

\newenvironment{geoffitemize}
{
	\begin{itemize}
}
{ 
	\end{itemize}
}

\fi

\begin{document}

\ifnum\anonymize=1
\newcommand{\SCS}{Hedgehog}
\else
\newcommand{\SCS}{Groundhog}
\fi

%don't want date printed
\date{}

\title{\Large \bf \SCS{}: Linearly-Scalable Smart Contracting via Commutative Transaction Semantics}

\if\anonymize1

\author{
	{\rm Submission 16}
}

\else
\author{
{\rm Geoffrey Ramseyer}\\
Stanford University
\and
{\rm David Mazières}\\
Stanford University
}
\fi

\newcommand\XXX[1]{\textcolor{red}{\textbf{XXX:} #1}}

%-------------------------------------------------------------------------------
\begin{abstract}

\SCS{} is a novel design for a smart contract 
execution engine based around concurrent execution of blocks of
transactions.  Unlike prior work, transactions within a block in \SCS{}
are not ordered relative to one another.  Instead, our key design insights
are first, to design a set of commutative semantics that lets the \SCS{} 
runtime deterministically resolve concurrent accesses to shared data.
Second, some storage accesses (such as withdrawing money from an account)
conflict irresolvably;
\SCS{} therefore enforces validity constraints on persistent storage accesses 
via a reserve-commit process.

These two ideas give \SCS{} a set of semantics that, while not as 
powerful as traditional sequential semantics,
are flexible enough to implement a wide variety of important applications,
and are strictly more powerful than the semantics used in some production blockchains today.

Unlike prior smart contract systems, transactions throughput never
suffers from contention between transactions.  Using 96 CPU
cores, \SCS{} can process more than half a million payment transactions per
second, whether between 10M accounts or just 2.

\end{abstract}

\settopmatter{printfolios=true,printacmref=false}
\maketitle
\pagestyle{plain}

\section{Introduction}

Blockchains offer many attractive features such as resiliency,
transparency, and, most importantly, the ability to transact atomically
across mutually distrustful parties.  A blockchain could potentially
act as a global cross-service coordinator, allowing atomic
transactions across arbitrary databases that were not explicitly
built to work together.   
Smart-contracts provide
extensibility as revolutionary as JavaScript in the browser or apps on
phones, and have had a huge multiplier effect on innovation by
attracting more developers with self-serve deployment.
But are we
ever going to see blockchains used for significant payment volume and throughput?  
What would
it take for a central bank or consortium of commercial banks to adopt
a blockchain for fiat currency payments?

One realistic path for mainstream adoption is to implement
a wholesale central bank digital currency (wCBDC\@).  A wCBDC would
connect banks and other financial institutions to each other and to 
power-users and
developers, which would lower the bar for innovation without the need
for average citizens to manage crypto wallets or the threat of
liquidity leaving our fractional reserve banking system.
Unfortunately, existing blockchains offer inadequate payment-per-second rates
for such a critical application.  Worse than actual
throughput numbers, blockchains have poor asymptotic scaling---they
derive limited benefit from more CPU cores, particularly in the worst
case of high contention (\cite{gelashvili2023block,garamvolgyi2022utilizing,saraph2019empirical}).
Designs that divide state between ``shards'' \cite{skychain,polkadot,nearsharding} or ``roll-ups'' 
\cite{poon2016bitcoin,kalodner2018arbitrum,poon2017plasma,loopringdesigndoc} %zkrollup,polygon,optimisticrollup,goel2020continuous
 accelerate only particular
payment patterns and complicate any cross-shard transaction.

A core challenge is that most blockchains are structured as
deterministic replicated state machines, which in existing systems has
hindered expressiveness, the use of hardware parallelism, or both.
Outside the blockchain context, transaction processing typically
exploits hardware parallelism by using locks for serialization
in case of transaction contention.  

Unfortunately, if transaction results depend on
ordering (e.g., the second of two \$10 payments from an account with
only \$15 must fail),
then inter-thread scheduling will introduce non-determinism in
transaction results, 
replicas will diverge, and the blockchain will fail.

Specifically, a state machine implementing a wCBDC must display at least
the following properties.

\begin{geoffitemize}

    \item \textbf{Deterministic Operation} 
    Replicas of the state machine
    must stay in consensus, and furthermore, past transactions
    must be replayable to enable auditing of historical data.

    \item \textbf{Flexibility} 
    The state machine should support executing arbitrary, untrusted programs,
    so as to support as-yet-uninvented applications and protocols.

    \item \textbf{Scalability}  Any significant piece of financial infrastructure
    is extremely difficult to modify once deployed.  A wCBDC design
    should be able to scale to transaction volumes far beyond those seen today.

    \item \textbf{Robustness}
    No malicious smart contract, nor any bug in any contract, should be able to degrade
    system performance or reduce transaction throughput.

   % and should maintain consistent throughput in the face of worst-case (potentially adversarial) 
   % inputs. %mention that worst-case inputs have taken down existing systems/caused negative externalities

\end{geoffitemize}

\paragraph{Our Contribution: \SCS{}}

This paper presents \SCS{}, a deterministic smart-contract wCBDC
ledger designed to achieve high and scalable worst-case performance.
\SCS{} introduces a new point in the design space,
extending the famous replicated state machine paradigm \cite{castro:bfs}:
a sequence of blocks of unordered, concurrent transactions.
\SCS{}'s semantics ensure that the execution order of transactions
in a block cannot ever affect the result of applying the whole block
(which enables efficient parallelization ~\cite{clements2015scalable}).

%An \SCS{} implementation executes 
%To execute a block of unordered
%transactions deterministically, we must ensure they commute---i.e.,
%have semantics unaffected by execution order (which enables efficient
%parallelization~\cite{clements2015scalable}). 
% \SCS{} achieves
%commutativity by executing everry
%transaction 

\SCS{} creates these semantics by adjusting the way in which
transactions read and write to ledger state (a key-value store).
Every transaction
in a particular block reads from
the identical snapshot of the ledger state at the start of the block.
It collects and defers transaction side-effects, each of which
has a structured type.
Most side-effects are
commutative (e.ġ., increment an integer) and can be combined
independently in the local memory of each CPU before being applied to
the global ledger state at the end of execution.

Of course, smart contracts do sometimes need sequencing---e.g., to
avoid double-spending a token balance.  \SCS{} shifts the burden of
synchronization from the ledger to smart contracts, which must ensure
that
transactions that require sequential execution
cannot be included in the same block.  More generally, contracts
can dynamically provide to \SCS{} a set of constraints on ledger state,
and replicas, when proposing a block of transactions, ensure that these 
constraints are always satisfied.  A reserve-commit process
makes block assembly efficient and scalable.

A few
simple abstractions in the API make this much simpler than it sounds.
\SCS{} provides non-negative integers to which
contracts can add and subtract.
%  Conceptually,
%one can think of the subtractions happening at the beginning of the
%block and the additions at the end.  
A set of transactions
\emph{conflicts} and cannot be included in the same block if the
subtractions would ever bring the integer below zero.
These non-negative integers are a perfect primitive to represent account balances
and semaphores.  \SCS{} also provides simple byte strings (where two writes of different
values conflict) and ordered sets
(where duplicate insertions conflict).  
%Non-negative integers
%(and other objects) act as cohorts in a 2-phase commit protocol,
%allowing a block proposer to efficiently assemble a batch of transactions
%that does not violate any constraint.

%Non-negative
%integers are a perfect primitive to represent account balances and
%mutexes.  Two other abstractions provided are simple byte strings
%(where two writes of different values conflict) and ordered sets
%(where duplicate insertions conflict).  
%Ordered sets allow smart
%contracts to implement transaction replay prevention and auctions in a
%natural way.

Although
these snapshot semantics are not as general as strictly
serializable semantics (as 
\SCS{} cannot execute arbitrary ordered transactions in the same block),
we
are nonetheless able to implement all of the popular smart contract
types we tried, including payments, auctions, an automated market
maker, and a partially-collateralized algorithmic money-market protocol.
Implementing these applications in \SCS{} typically requires only small API changes.
As such, we expect most users not to use \SCS{}'s APIs directly, but rather,
to write contracts leveraging higher-level application-specific APIs.

\SCS{}'s semantics are strictly more powerful than the 
asynchronous message-passing architectures implemented in some 
large-scale blockchains \cite{nearcrosscontract,cosmwasmactor}. 
Furthermore, \SCS{} does not require sharding or otherwise subdividing
the blockchain's state to achieve its parallel execution, does not require
precomputation of transaction writesets, and does not rely
on any manual analysis of untrusted smart contracts.

As an additional benefit compared to other blockchains,
\SCS{} eliminates the ability for block producers to extract value
by ordering the transactions within a block, a widespread phenomenon in current systems \cite{daian2019flash,qin2022quantifying} %ELASTIC citation
(although a malicious replica can still delay transactions). 
Our evaluation of \SCS{} shows
that it scales nearly linearly to 96 cores (the most we have available) no matter
the level of contention between transactions,
and handles more than half a million payments per second between 10M accounts.
%Smart contracts in \SCS{} are compiled to WebAssembly modules.

% Groundhog benefits: performance, scalability, no front running

%% Front-running?

%% UTXOs suck: many reasons
%% ---
%% Note that many systems based on so-called ``unspent transaction outputs'' (UTXOs) already suffer from this type of limitation (e.g. \cite{sundaeswap}),
%% simply because (without external infrastructure) one user cannot spend a UTXO until they learn of its existence, which happens when it is broadcast in a block proposal.
%% ---

\section{Architecture Overview}
\label{sec:architecture}

\SCS{} is a deterministic, scalable execution engine for a replicated state machine,
with the following components:

\begin{geoffitemize}
  \item \textit{Smart contracts}, implemented as WebAssembly modules, are deployed at 32-byte
  \textit{contract addresses}. 
  \item Each smart contract has a key-value store, using 32-byte keys 
  (so overall system state is a key-value store, where keys are 64-byte \textit{addresses}).
  Values are \textit{typed objects} (\S \ref{sec:storage_types}).
  %\item Each contract may possess some local storage, which consists of \textit{typed objects} (\S\ref{sec:objects}).
  %Each object is referenced by a 32-byte identifier in the contract's namespace (so each object has a unique 64-byte identifier).
  \item Each \textit{transaction} is a function call on a smart contract, specified by a contract address, a method,
  input (bytes) for the function, and some metadata.
  Contracts may call into other contracts.
\end{geoffitemize}

What makes \SCS{} unique is the way that it processes blocks of transactions (Algorithm \ref{alg:basic_exec}).  
Instead of executing individual transactions sequentially,
\SCS{} executes blocks of transactions \textit{concurrently}.
Semantically, there is no ordering relation between transactions in the same block.
Instead, \SCS{} makes a snapshot of the entire state machine before processing any transactions in a block.
When smart contracts read from the key-value store,
they read the data in this snapshot.

Smart contracts modify the state machine through \textit{typed modifications}; that is, the output of a transaction
is a list of modifications $\lbrace (k_1, \delta_1),...,(k_l, \delta_l)\rbrace$.
The \SCS{} runtime groups the modifications produced within a block
by address. 
Then, for each address individually, \SCS{} applies, all at once, all of the modifications to that address 
(Algorithm \ref{alg:basic_exec}, lines \ref{line:start_inner_apply_for}-\ref{line:end_inner_apply_for}).

\newlength{\textfloatsepsave} 
\setlength{\textfloatsepsave}{\textfloatsep} 
\setlength{\textfloatsep}{0pt}

\begin{algorithm}[t]
\caption{\SCS{}'s block execution (simplified)}
\label{alg:basic_exec}
\KwIn{The next block of transactions $\lbrace \tx_1,...,\tx_n \rbrace$}
\KwIn{The system state (key-value store) $S=\lbrace (k_1,v_1),...,(k_m, v_m) \rbrace$}
$\hat{S}\gets S$ \tcc*{Take a snapshot of $S$}
\ForEach{Transaction $\tx_i$}{
  Execute $\tx_i$.  Reads observe $\hat{S}$. \\
  $\tx_i$ produces a list of modifications to key-value pairs
  $((k_{i_1}, \delta_{i,1}),...,(k_{i_l}, \delta_{i,l}))$.
}
Group modifications by key \\
\ForEach{$(k,v)$ in $S$}
{
  $v^\prime \gets v$\\
  \ForEach{$\delta_{i,j}$ applied to $k$ by some transaction $\tx_i$ \label{line:start_inner_apply_for}}{
    $v^\prime \gets \textsf{apply}(v^\prime, \delta_{i,j})$ 
  }\label{line:end_inner_apply_for}
  Assert $v^\prime$ satisifies \textit{validity constraints} \label{line:assertion}\\
  Replace $v$ by $v^\prime$ in $S$
} \label{line:end_apply_for}
\KwOut{The updated state $S$}
\end{algorithm}

 \setlength{\textfloatsep}{\textfloatsepsave}

The contract and storage addressing scheme
and cross-contract call semantics borrow from the standard design patterns used in public blockchains.% (e.g. \cite{wood2014ethereum,solanaaccount}).

\subsection{Managing Concurrent Writes}

\SCS{}'s model is equivalent to the standard sequential consistency model
if each block contains only one transaction.  In such a setting,
a transaction would modify the state machine by simply writing bytestrings to particular addresses.

However, when a block contains many transactions (as in \SCS{}'s target parameter regime),
multiple transactions in the same block may modify a particular key.  We call such modifications \textit{concurrent}
(even if an implementation does not execute them at the same wall-clock time).

A core challenge in \SCS{}'s design is in managing these concurrent writes.  \SCS{} relies on two core ideas.
First, we add type information to values in the key-value store,
and transactions change these values with \textit{commutative}, typed modifications.
This information lets the \SCS{} runtime automatically resolve many concurrent writes,
and commutativity crucially ensures that
the order in which a replica executes this resolution process does not affect the result.
For example, a value could have an integer type, which transactions modify through addition and subtraction.
We defer the details of \SCS{}'s types to \S \ref{sec:storage_types}.
Concretely, commutativity means that the loop in lines ~\ref{line:start_inner_apply_for}-\ref{line:end_inner_apply_for} 
of Algorithm \ref{alg:basic_exec} produces a deterministic result, regardless of the iteration order.

And second, applications need constraints on their state,
and not all concurrent modifications can be resolved.
For example, a digital cash system might require that every user's
account balance be nonnegative, and concurrent writes that, e.g., register two different public keys
inside a contract cannot be resolved.
Therefore,
\SCS{}
also enforces type-specific \textit{constraints} 
on the total set of concurrent modifications to a key.  
We defer description of specific constraints to \S \ref{sec:storage_types}.

\SCS{} implements these constraints as a set of assertions (Algorithm ~\ref{alg:basic_exec}, line ~\ref{line:assertion}).
\SCS{} rejects a block of transactions if there are any concurrent modifications which it cannot resolve, or sets of modifications which
violate a type-specific constraint.
For example, in a digital currency system with sequential semantics,
if two transactions spend the same coin, the first transaction would succeed
while the second would fail; instead, \SCS{} ensures
that a block contains no such conflicting pairs.
Fortunately,
a reserve-commit process makes proposing a new blocks of transactions
nearly as efficient as executing a fixed block of transactions.
When proposing a new block, a replica chooses to include a transaction
only when it knows that it will not conflict with any transaction
already in the block.  
Algorithm \ref{alg:parallel_exec} gives an overview.

The crucial property of the type-specific $\textsf{reserve}()$ methods
is that a $\textsf{reserve}()$ call can be safely rolled back anytime before a corresponding
$\textsf{commit}()$ call.  This property implies that aborted transactions
cannot cause Algorithm \ref{alg:parallel_exec} to improperly output a block
containing unresolvable, concurrent modifications to a key.
This property is what allows a \SCS{} replica to efficiently
propose new blocks of transactions
in the presence of malicious users and smart contracts.
\footnote{
  Each WASM runtime is metered, to guard against nonterminating contracts.
}
We defer type-specific details of this process to \S \ref{sec:lockfree}.

\setlength{\textfloatsepsave}{\textfloatsep} 
\setlength{\textfloatsep}{0pt}

\SetKw{Outerloop}{continue outer loop}

\begin{algorithm}[t]
\caption{\SCS{}'s block proposal}
\label{alg:parallel_exec}
\KwIn{A stream of new transactions $T=\lbrace \tx_1,\tx_2,... \rbrace$}
\KwIn{State $S=\lbrace (k_1,v_1),...,(k_m, v_m) \rbrace$}
$\hat{S} \gets S$ \tcc*{Take a snapshot of $S$}
$B\gets \emptyset$\\
\Repeat{$B$ is large enough}
{
  Draw new $\tx_i$ from $T$ \\
  Execute $\tx_i$.  Reads observe $\hat{S}$. \label{line:tentative}\\
  $\tx_i$ produces a list of modifications to key-value pairs
    $((k_{i_1}, \delta_{i,1}),...,(k_{i_l}, \delta_{i,l}))$. \\
  \label{line:start_rc_loop}
  \ForEach{$\delta_{i,j}$ modifying $(k_{i_j}, v) \in S$}{
    \If{$\neg\textsf{reserve}(v, \delta_{i,j})$}{
      Abort $\tx_i$ \tcc*{undo $\textsf{reserve()}$ calls}
      \Outerloop
    }
  }
  \ForEach{$\delta_{i,j}$ modifying $(k_{i_j}, v) \in S$}
  {
    $\textsf{commit}(v, \delta_{i,j})$
  }
  $B\gets B\cup \lbrace \tx_i\rbrace$
}
\KwOut{The updated state $S$}
\KwOut{Proposed block $B$}
\end{algorithm}

\setlength{\textfloatsep}{\textfloatsepsave}

In our implementation,
both proposing a new block and executing a given block
use the same code, based on Algorithm~\ref{alg:parallel_exec}.
When assembling a new block, a replica
can abort candidate transactions nondeterministically,
but when (deterministically) executing a given block,
an unresolvable concurrency conflict requires aborting the entire block.
%Say something about nondeterminitic proposal, throw out txs
% bad block, throw out whole block
The overhead of block proposal
is the possibility of tentatively executing transactions (line~\ref{line:tentative})
that may be aborted.

\subsection{Deployment Context}
\label{sec:deployment_context}

We envision \SCS{} deployed within a large-scale digital currency system,
where each replica is operated by a central bank or a large financial institution---institutions
with the resources to deploy high-performance systems with many CPUs.
These replicas would be connected with some consensus protocol.  \SCS{} is agnostic to the choice of protocol,
but is designed around a context where a leader periodically (once per second, perhaps)
proposes a block of transactions, such as in SCP~\cite{lokhava:stellar},
HotStuff~\cite{yin:hotstuff}, or BA$\star$~\cite{gilad:algorand}.
Proposed blocks would then be checked for correctness 
(that is, the absence of unresolvable, concurrent modifications or constraint violations) by other replicas.

Repeated proposal of invalid blocks is a potential denial of service vector for a
system using \SCS{}.
In this context, we imagine that replica operators can be sanctioned
if they repeatedly propose invalid blocks.
Malicious users may cause a block proposer
to waste CPU time,
but cannot cause an honest replica to propose an incorrect block.

\subsection{Design Properties}

Our design for \SCS{} meets our requirements for the execution engine of a digital currency system.

\paragraph{Deterministic Operation}

Transactions modify keys 
with \textit{commutative} typed modifications (\S \ref{sec:storage_types}).
This lets \SCS{} resolve concurrent modifications to an address deterministically.
And because each transaction executes in isolation (reading from a snapshot), each transaction executes deterministically.

\paragraph{Parallel Proposal and Execution}

The main loop of Algorithm \ref{alg:parallel_exec} is arbitrarily parallelizable,
with almost no data dependencies between loop iterations.
Cross-thread write synchronization is limited to the $\textsf{reserve()}$ and $\textsf{commit}()$ methods
of individual values in the key-value store.  These are implemented
with hardware atomic memory instructions (\S \ref{sec:lockfree}).

After iterating over all transactions in a block, \SCS{} iterates over all modified objects
in the key-value store
to ready the state snapshot for the next block,
and to produce appropriate state commitment (i.e. root hashes) useful for a digital currency application.
This is again (nearly) arbitrarily parallelizable (\S \ref{sec:lockfree}).

To emphasize, \SCS{} executes in parallel both when proposing a block (as a leader)
and when executing a proposal from another replica.
When the reserve-commit process of Algorithm~\ref{alg:parallel_exec}
aborts a transaction,
a block proposer drops the offending transaction from its proposal,
while other replicas raise an error and roll back the entire block.
There is never a fallback to sequential execution.

\iffalse

The semantics of \SCS{} are designed to enable effective parallelization of block proposal and execution
while minimizing inter-thread synchronization overhead.  Because each transaction in a block reads from 
a snapshot of state, and never reads data written by transactions in the same block, an implementation
can largely avoid cross-thread synchronization.  The only coordination required is for the maintenance
of object constraints, which our implementation achieves using only hardware-level atomic memory instructions
(\S \ref{sec:proposal}).  After iterating over all transactions in a block, \SCS{} iterates (again, in parallel) over
every modified object to produce the state snapshot for the next block.

To emphasize, \SCS{}'s parallelism is guaranteed for all replicas, both when proposing a block (as a leader)
and when executing a proposal from another replica (as a follower).  In fact, leaders and followers execute nearly the same code path.
When the 2-phase locking algorithm of \S \ref{sec:proposal}
identifies a conflict,
a leader drops the offending transaction from the proposal it is assembling,
while a follower raises an error and rolls back the block.
There is never a fallback to sequential execution.

\fi

\paragraph{Semantic Flexibility}

The smart contracts in \SCS{} are arbitrary WebAssembly modules, and do not require validation
or static analysis.
The semantics of how these modules interact with the state machine are different from standard strictly serializable semantics.
However, as described in Fig. \ref{fig:contracts},
we are able to implement many widely-used applications
in these semantics, with minimal changes to the APIs these applications give to downstream smart contracts.
Furthermore, these semantics are strictly more powerful than those based on message-passing actors (\S \ref{sec:sequencers}) used in
some blockchains deployed in production today.

\begin{figure*}
\begin{center}

\begin{tabularx}{\textwidth} { 
  | >{\raggedright\arraybackslash}X
  | >{\centering\arraybackslash}c%\hsize=0.5\hsize}X
  | >{\centering\arraybackslash}X
  | >{\centering\arraybackslash}c
  | }
\hline
Application & Implementable & API Changes & Atomic Composability (\S \ref{sec:sequencers}) \\
\hline
\hline
Tokens (ERC20 Standard~\cite{eip20}, \S \ref{sec:token}) 
	& \contracttableyes & \{inc,dec\}reaseAllowance\cite{erc20openzeppelin,erc20attack} & \contracttableyes \\
\hline
k-th Price Auction & \contracttableyes & None & \contracttableyes \\
\hline
%Automated Market-Maker (Uniswap v2~\cite{uniswapv2}, \S \ref{sec:cfmm})
Automated Market-Maker (\S \ref{sec:cfmm})
	& \contracttableyes & None & \contracttableno \\
\hline
Money Market (Compound v3 ~\cite{compoundv3}, \S \ref{sec:compound}) &
\contracttableyes & Separate Borrow and Supply Accounts & \contracttableyes \\
\hline
Payment Channels \cite{poon2016bitcoin} & \contracttableyes & None & \contracttableyes \\
\hline

\end{tabularx}

\end{center}
\caption{A sample of applications implementable in \SCS{}, and the changes required by \SCS{}'s model.
\label{fig:contracts}
}
\end{figure*}

\paragraph{Robustness to Malicious Contracts}

We emphasize that contracts in \SCS{} can be untrusted, arbitrary scripts subject to no
static analysis or manual review.
Although of course smart contracts deployed on \SCS{} may contain bugs,
malicious or erroneous code cannot break down the operation of \SCS{}.
A transaction may waste compute resources, but because each transaction
executes in isolation against a static snapshot of state, a malicious transaction
cannot manipulate the execution engine to cause, for example, worst-case performance of a
concurrency control algorithm.  And as discussed in \S \ref{sec:deployment_context},
malicious contracts cannot cause incorrect operation of any replica.

\TODO{What about failing txs that don't get added in leader?}
%\SCS{} can charge each transaction for the resources
%it consumes, although \SCS{} is vulnerable to the same challenges as other blockchains
%with regard to accuracy of resource pricing \cite{perez2019broken}.

\subsection{Design Motivation}

Conceptually, in one block,
\SCS{} concurrently executes a collection of user-defined applications.
Instead of trying to accelerate these unknown applications, \SCS{} leaves it to users to build 
support for concurrency and implement
their own locks (\S \ref{sec:locks}).  
An application that relies on strictly sequential semantics can acquire a lock on its state, but others might allow fine-grained locking 
or may not require locks.  This isolates the performance (in terms of transaction throughput) of high-throughput applications
from that of incorrect, slow, or malicious ones.

Our claim is not that every application can be reimagined as one that admits concurrent updates.
Rather, we observe that many important financial applications can 
support at least some level of concurrent execution.  And \S \ref{sec:sequencers}
gives a worst-case fallback option for contracts that need to interact with
other contracts but require sequential semantics (such as constant function market makers, \S\ref{sec:cfmm}).

\section{Storage Semantics}
\label{sec:storage_types}

We describe here the specific types that values in \SCS{}'s
take,
and the API through which transactions modify these values.

We implement this API
by providing our WebAssembly interpreter
a set of specific methods to access the key-value store.
We additionally provide a basic set of operations
common to many blockchain contexts, such as 
$get\_block\_number()$, $get\_caller\_address()$, \penalty 0 $get\_self\_address()$,
\penalty 0 and a few cryptographic primitives.

\def\strset{\textsf{string\_set}}
\def\strget{\textsf{string\_get}}

\def\intsetadd{\textsf{int64\_set\_add}}
\def\intget{\textsf{int64\_get}}
\def\intadd{\textsf{int64\_add}}

\def\hsins{\textsf{set\_insert}}
\def\hsclr{\textsf{set\_clear}}
\def\hslim{\textsf{set\_limit\_increase}}

\def\hsgetindex{\textsf{set\_get\_index}}
\def\hslookup{\textsf{set\_lookup}}

\subsection{Bytestrings}

A value could be an uninterpreted string of bytes. 
Most applications need some configuration data,
for example,
and in fact most public blockchains implement all stored state in this manner.
\sloppy Contracts access a bytestring through methods $\strget()\rightarrow string$ and $\strset(string)$.
Two concurrent calls $\strset(x)$ and $\strset(y)$ conflict if and only if $x\neq y$.

\subsection{Nonnegative Integers}
\label{sec:nnint}

\sloppy \SCS{} implements a ``nonnegative integer'' type.
As an example, this type implements account balances (\S \ref{sec:token}),
as well as semaphores and linear constraints (\S \ref{sec:locks}).
%This type effectively implements account balances,
%but also semaphores and arbitrary linear constraints (\S \ref{sec:locks}).

Contracts access these integers through methods $\intget()\rightarrow int64$
and $\intsetadd(int64, int64)$.
A call to $\intsetadd(x, y)$ sets the value of the integer to $x$,
and then adds $y$ to it.
Two concurrent calls $\intsetadd(x_1,y_1)$ and $\intsetadd(x_2,y_2)$ conflict
if $x_1\neq x_2$, but if $x_1=x_2$, then the overall
 result of applying both modifications concurrently
 is equivalent to applying $\intsetadd(x_1, y_1 + y_2)$.

This interface allows concurrent modification of the same integer.
Consider the case of an account balance,
and suppose several transactions concurrently debit or credit a balance by amounts $y_1,...,y_k$.
The net result of applying these transactions (assuming sufficient funds)
in a sequential system would be the initial account balance, plus $\sum_i y_i$.
In \SCS{}, this transaction sequence would correspond to concurrent calls to
 $\intsetadd(\intget(), \sum_i y_i)$ on the account.  These modifications, when applied all together,
would set the account balance to its initial value, plus $\sum_i y_i$.
The \SCS{} API also provides a $\intadd(int64)$ method for convenience,
which is equivalent to $\intsetadd(\intget(), int64)$.

\paragraph{Constraint: Nonnegativity}
\SCS{} constrains these integers to be nonnegative.
This constraint is necessary for implementing account balances, for example.
Specifically, a set of concurrent calls $\lbrace \intsetadd(x, y_i)\rbrace$
violates this constraint if $x\geq 0$ but $x+\sum_i y_i <0$, or if $x<0$ and any $y_i < 0$.

In designing \SCS{}'s semantics, we choose to allow ``nonnegative integers'' to be explicitly set to a negative value
if a contract so desires
because we err on the side of flexibility, 
and this behavior appears useful in narrow cases (\S \ref{sec:compound} gives an explicit example).
If a contract never explicitly sets a negative value, then \SCS{} maintains the invariant that the
value will never be negative.

\subsection{Ordered Sets}
\label{sec:sets}

The last type that \SCS{} implements is an ``ordered set.''
This type implements the abstraction of a collection of (unique) objects,
sorted according to a user-specified index.

Elements of these sets are tuples $(\text{tag},\text{hash})$, where the tag is a 64-bit unsigned
integer
and the hash is a 32-byte string.
When \SCS{} makes a snapshot of the key-value store,
it sorts each of these sets by tag, in ascending order and breaks ties by comparing hashes.
Smart contracts can call $\textsf{set\_get\_index}(uint32)\rightarrow string$ to read the $n$th element of the set,
or access with $\textsf{set\_lookup}(uint64)$ the first element with a tag at least the queried threshold.
Sets of more complex and larger objects can be implemented by interpreting each hash in the set as an address
in a contract's local storage that points to some other object.

Smart contracts can $\hsins(uint64, string)$ an element to the set, or $\hsclr(uint64)$ all of the elements in the set
with tags less than a threshold.
These two actions do not obviously commute; logically, \SCS{} applies all insertions in a block, 
before applying any of the concurrent clear operations, thereby making the overall process of resolving concurrent accesses
commutative.

\paragraph{Constraint: Unique Hashes}

\SCS{} requires each of the 32-byte strings in the set to be unique.
That is, two concurrent calls $\hsins(t_1, h)$ and $\hsins(t_2, h)$ conflict
(even with different tags).

\paragraph{Constraint: Size Limit}
\SCS{} also imposes a size limit on each ordered set,
to prevent malicious transactions from imposing excessive costs on an implementation
via arbitrarily large sets, as well as to encourage garbage-collection of stale elements.
By default, a set's size is limited to 64 elements,
but \SCS{} allows this limit to be raised (to a maximum of 65535)
via a $\hslim(uint16)$ operation.  Limit increases beyond 65535 have no effect.

\subsection{Creation and Deletion}

Values are created implicitly when they are first written to by a transaction.
A new value is initialized to a default,
depending on the type of value implied by the access.
Bytestrings are initialized as empty strings,
integers are initialized as 0,
and sets are initialized as empty.
Naturally, two transactions that concurrently create different value types at the same address
conflict.

Transactions can explicitly $\textsf{delete}()$ values.  Deletions are applied after all other concurrent writes
to a value, and two concurrent deletions do not conflict.

\section{Lock-free Reserve-Commit}
\label{sec:lockfree}

\def\reserve{\textsf{reserve}}
\def\commit{\textsf{commit}}

\SCS{} parallelizes the main loop of Algorithm \ref{alg:parallel_exec}.
The key to high performance of this loop (and near-linear scalability)
is in minimizing cross-thread synchronization.
\SCS{} uses a reserve-commit system,
which isolates concurrent memory accesses to the level of individual
values in the key-value store,
and can be implemented efficiently with just a few hardware atomic memory instructions.

The required invariant is that any $\reserve()$ must be safely
reversible if the transaction aborts.  Without this invariant, one
might implement a nonnegative integer with a single atomic counter,
where $\reserve()$ for $\intadd(x)$ just adds $x$ to the counter.
However, undoing such a $\reserve()$ of $+x$ would mean subtracting
$x$ from the counter which, depending on actions in other threads,
could drive the counter below $0$.

On any $\reserve()$ call, $\SCS{}$ first checks
for (or sets) the parameters that must be the same 
across concurrent modifications of that value.
This includes the type of the value, as well as type-specific 
data---the 
string for a $\strset()$, and the base value in $\intsetadd()$.
These parameters, along with some metadata (the number of threads
that expect these particular parameters, which allows a thread
to determine whether to clear the parameters upon rolling back a $\reserve()$)
are tracked with an atomic swap on a pointer to an immutable copy of the parameters.
No thread knows precisely when a copy of the parameters can be freed---another thread might
be concurrently reading the data---%
so \SCS{} defers garbage collection to the end of processing a block (which also avoids the ABA problem \cite{dechev2010understanding}).

\paragraph{Nonnegative Integers}

\sloppy We implement a nonnegative integer with \emph{two} atomic counters;
one tracks the total amount added to the integer,
and the other tracks the total amount subtracted from it.
A $\reserve()$ call corresponding to an $\intsetadd(x,y)$ operation
adds to the appropriate counter, and checks (in the case of a subtraction)
whether the total amount in the subtraction counter would exceed the base value ($x$) of the integer.

Undoing a $\reserve()$ call means subtracting from the appropriate counter.
Note that this is always safe; $\reserve()$ maintains the invariant
that the subtraction counter is less than the base value ($x$), so the value of the integer
at the end of the block---that is, the base value, less the subtraction counter, plus the add counter---is always nonnegative.

\paragraph{Ordered Sets}

We implement ordered sets with an atomic hashmap.  The size limit sidesteps the implementation
complexity of resizing the hashmap in a lockfree manner.
Elements of this hashmap are pointers to the actual 32-byte hashes.  Again, because a thread cannot easily know
whether another thread is accessing a given 32-byte hash object,
garbage collection is deferred to the end of the block.
We allocate hashes from preallocated buffers, allowing our implementation to use 32-bit pointers in the hashmap,
and garbage collection is just resetting a per-thread offset index.

$\reserve()$ for $\hsclr()$ is a no-op, with the actual work deferred to $\commit()$.
This lets our implementation track the maximum committed clear threshold with a single integer.

\section{Implementation}
\label{sec:impl_short}

Our implementation of \SCS{} consists of approximately 30k lines of C++.  
Our contract SDK is written in C++, but \SCS{} itself is agnostic to the
choice of source language.
Contracts are compiled to WebAssembly and then executed using the Wasm3 interpreter \cite{wasm3}.
Before executing a contract,
\SCS{} instruments the WebAssembly module with duration metering (``gas metering'') 
via \cite{wasminstrument}.

\SCS{} and our example contracts are available at \githuburl{}.

\subsection{Data Structures}
\label{sec:data_structures}

Our implementation provides the features that would be important to a real-world deployment of \SCS{},
although a complete digital currency system would include \SCS{} as one part of a larger system.
Persistent state lives in Merkle-Patricia tries; this is required for
an implementation to produce short proofs of state or of a transaction's execution.
Our implementation includes transaction runtime metering but not a default unit of money.

After executing a batch of transactions,
\SCS{} must iterate over all modified elements of the key-value store to compute a new state snapshot for the next block.
To make this efficient, it builds another trie while executing a block
that logs which keys are modified.
The prefixes of this trie are the same as those of the (much larger) 
main key-value store,
which means iterating over this trie enables efficient iteration over
the modified elements of the main key-value store.
\SCS{} also builds a trie logging all of the transactions in the batch.
Finally, \SCS{} recomputes the root hashes of each of these tries.

Our implementation stores these tries in memory, and writes data to persistent storage
as necessary.  
\SCS{} specifically logs every new block of transactions
and every changed node (i.e., at most once per block) in the Merkle-Patricia trie
that represents the main key-value store.
Whenever a node in this trie is logged to persistent storage,
\SCS{} assigns it a new, unique identifier.
Records of non-leaf nodes also store the logged identifiers
of their child nodes, which suffices to recover the trie state
from these logs.
Due to hardware constraints (lack of storage drives)
some of the experiments of \S \ref{sec:scalability} disable persistent storage.

%We use Merkle-Patricia tries, instead of a possibly faster data structure,
%because a real-world deployment of \SCS{} would likely want to produce short cryptographic
%proofs that a transaction was included in a batch, or that a user's balance is a particular value.
%The high transaction throughput rates for which we design \SCS{} inherently require that \SCS{} replicas 
%have many more compute resources than regular users, and certainly a smartphone is not likely to be able to operate
%a full copy of \SCS{} in realtime in the near future.  Users on mobile devices, however, might still like to verify
%that a transaction occurred, which requires the short proofs that Merkle tries enable.  

%Our goal is to implement the features important to a real-world deployment that require direct integration with \SCS{}.
%A deployment of a digital currency would include \SCS{} as one part of a much larger system.  Our implementation
%stores all data in memory, but a deployment would log transactions to persistent storage. 
%\SCS{} builds a trie capturing all of the modified storage locations
%to enable an efficient implementation to identify all locations that must be rewritten to persistent storage.
%while persistent logging may not need to
% be on the critical path, an efficient implementation would need to quickly identify all of the storage locations modified in a batch.

%A deployment could implement a fee token using a standard token contract (\S\ref{sec:token})
%deployed at a hard-coded location.

%\input{commutative_semantics}

\section{Evaluation: Semantic Flexibility}

We demonstrate the flexibility \SCS{}'s semantics by discussing how we
implemented a variety of applications commonly used on existing
blockchains.

\subsection{Tokens}
\label{sec:token}

\SCS{}'s semantics implement a fungible token, where every user
has an account balance.  We use a variant of the widely-used ERC20 standard token API \cite{eip20}.
Balances are represented as nonnegative integers (\S \ref{sec:nnint}).

We make an important change to the public token interface.
In the ERC20 standard, one contract can explicitly $\textsf{approve}(x)$
 another to transfer $x$ number of tokens on its behalf.
Instead, \SCS{} manipulates these approval amounts by adding and subtracing to them.

This interface not only enables an efficient implementation in \SCS{}
(with nonnegative integers) but also avoids a double-spend attack in
existing blockchains \cite{erc20attack}:  a malicious contract could
consume its allowance right before a user changes the authorized
amount, effectively gaining access to more of the user's balance than
the user intended to authorize.  Open-source implemenations of this
API in public blockchains \cite{erc20openzeppelin} recommend this API
change, even though public blockchains use strictly ssequential
semantics.

A token contract suffices to let \SCS{} implement 
payment channels \cite{poon2016bitcoin}, atomic token swaps,
and some decentralized exchanges that perform offer matching off-chain \cite{warren20170x}.  We
have implemented a sample of these applications.

%There is a wide class of applications for which even a minimal commutative token implementation suffices to enable a commutative implementation
%of the application.  This class includes payment channels \cite{poon2016bitcoin}, atomic token swaps, and 
%the subclass decentralized exchanges, such as 0x \cite{warren20170x}, that perform offer matching logic off-chain.  
%We have implemented a sample of each of these applications.

Prior work analyzed Ethereum's history \cite{garamvolgyi2022utilizing} and found that, under sequential semantics,
the majority of access patterns that prevent optimisitic parallel execution result from integer counters
and token balances.

\subsection{Locks and Linear Constraints}
\label{sec:locks}

An application may require that an action is only performed once, or that the data that a transaction reads is not concurrently modified by
another transaction.
Our nonnegative integers (\S \ref{sec:nnint})
exactly implements such a lock, through a call to $\intsetadd(1, -1)$.
Only one transaction can concurrently execute that operation on a particular address.
Multi-reader locks could be implemented as a $\intsetadd(0, 0)$ call.
One limitation of \SCS{} is that it lacks a mechanism for prioritizing transactions
contending on a lock.

This primitive also implements arbitrary linear constraints on a
contract's state.  Suppose, for example, that a contract requires that
certain variables $\lbrace x_1,\ldots,x_n\rbrace$ satisfy the
condition that $\sum_i a_ix_i \geq c$ for some constants $c,
a_1,\ldots, a_n$.  A contract would create an extra nonnegative
integer variable $y$ with value $(\sum_i a_ix_i - c)$ to represent
this constraint.  Changing a variable $x_j$ to $x_j^\prime$
additionally requires calling \sloppy $\intsetadd((\sum_i a_ix_i - c),
~(a_j(x_j^\prime-x_j))$ on $y$.
%$\intsetadd \left(\sum_i a_ix_i - c, \left(a_j(x_j^\prime-x_j)\right)\right)$.

\iffalse
This interface implements shared read-write locks as well.  Exclusive
access translates to a $set(1); add(-1)$ call, while shared access
translates to a $set(0)$ call.  That said, \SCS{} does not have a
mechanism for prioritizing which transactions succeed in the case of
contention on a lock.  A rudimentary solution would be for readers and
writers to divide access between even and odd block numbers.

\fi

\subsection{Replay Prevention}
\label{sec:replay}

A transaction can ensure that it only executes once by inserting a hash of itself into one of the ordered sets of 
\S \ref{sec:sets}.  This makes a transaction conflict with any concurrent or future invocation of itself.  
%An example mechanism
%is implemented in Fig. \ref{fig:replaymechanism}.  
We imagine that most user transactions would first call into a
``wallet'' contract that implements such a mechanism (as well as performing
authorization checks, such as signature validation).

%We imagine that users would deploy wallet contracts that
%would implement this feature on every transaction.

In this context, the tag on the ordered set enables garbage collection.
It would be impractical for a
ledger to store indefinitely (outside of high-latency archival storage) 
the hash of every transaction in its history; most blockchains implement or
propose some way to reclaim stale data~\cite{solanarent,ethstateexpiry}. %stellarexpirycap

Each transaction can implement an expiration time (block number), and (after checking that the current block number
is less than the expiration time)
use this time as
the tag on the hash in the ordered set.
It can then clear the set of all entires with tags less than the current block number, thus
garbage-collecting entries that are no longer necessary.

\iffalse

Therefore, a replay mechanism can include on each transaction an expiration time (batch number).
The mechanism can check whether the transaction has expired, and (if not) insert
the transaction's hash into a set, using as the tag the expiration batch number.
Clearing the set of all entries with tags less than the current batch number
thus garbage collects entries no longer necessary for replay prevention.

\fi

\iffalse
\begin{figure}

\begin{minted}[fontsize=\codefontsize]{c++}
void record_replay(uint64_t expiration_time) {
  sdk::hashset_insert(
    replay_cache_storage_key,
    sdk::get_invoked_hash(),
    expiration_time);

  uint64_t current_block_number 
    = sdk::get_block_number();
  if (expiration_time < current_block_number)
    abort();

  sdk::hashset_clear(
    replay_cache_storage_key, 
    current_block_number);  
}
\end{minted}

\caption{A function implementing a replay prevention mechanism
that garbage-collects expired transactions.
\label{fig:replaymechanism}
}

\end{figure}

\fi

\subsection{Auctions}
\label{sec:auctions}

Auctions form another core component of many financial applications.  Some protocols, such as MakerDAO
(the service behind the Dai stablecoin \cite{dai}), use auctions to liquidate collateral when a borrower's collateral to debt ratio falls.
%Many services issue indivisible assets (such as NFTs \cite{eip721}) using competitive auctions.  
Fierce contention of many transactions on a small set of auctions
can take down entire systems \cite{solanaoutage}.

The input to an auction is a set of bids, and the output is the winner and a winning price.
For example, the highest bidder often pays the
bid offered by the second-highest bidder.

We implement such an auction using the ordered sets of \S \ref{sec:sets}.  Users submit records of bids (an offered price, and the user's identity),
which are then hashed and inserted into a set.  The tag associated with the bid is the offered price.   After the auction closes,
% (either by an external
%action or after a certain time passes)
the bids in the set will sorted by their price---so a contract can easily compute the second highest bid.  
Users send the auction
contract capital
to cover their bids when creating a bid, and can refund this capital after the auction ends.  
To prevent a bid from being refunded twice,
users insert the hash of their bid into a second set after activating a refund.  

A second-price auction need only keep track of the top two bids.  Low
bids (that might otherwise fill a set to the maximum limit) can be
refunded and cleared from the sets (both the bid set and the refund
replay-prevention set) when necessary.  When receiving a new bid, the
contract should check to ensure that the new bid offers a price at
least as high as the current second-highest bid. 
% To avoid filling up
%the ordered set, the contract could also require each bidder to refund
%and delete a superseded bid, though we have not yet implemented such a
%feature.

\subsection{Money Market}
\label{sec:compound}

%We additionally implemented a lending and borrowing protocol.  
We additionally implemented the Compound protocol~\cite{compoundv3,leshner2019compound} (version 3) on Ethereum,
which manages several billion dollars worth of assets~\cite{compoundapp}.
Other money market protocols operate on similar principles.
Users supply or borrow a  ``base'' asset (in practice, a USD-pegged token).
Lenders and borrowers receive and pay interest, respectively.

Interest is implemented with compound interest index global variables. 
Base asset amounts are stored relative to amounts hypothetically deposited
at the beginning of the protocol and converted to present value using these indices
(so updating the indices implicitly updates every user's present balance).
Borrowing and lending occur at different rates,
and maintain separate indices.  

In the Ethereum implementation,
each transaction updates these indices,
based on the time elapsed since the last update.
Yet the observed ``time'' advances only once per block,
so these indices change once per block.
This kind of update fits naturally with \SCS{}'s
semantics, as concurrent writes of the same value to a bytestring do not conflict.

Compound overcollateralizes borrowing with deposits of other assets.  
Users can only borrow up to a collateralization limit.
\SCS{} implements this limit via a linear constraint on each user's state (\S \ref{sec:locks}).
Collateral values fluctuate and user balances change each block (due to interest),
so the value of this constraint may change from block to block.
Thus, every transaction on a user's data recomputes this limit, as the input to a $\intsetadd(limit, \cdot)$ operation), 
before adding or subtracting to it
(so two concurrent transactions from one user do not inherently conflict).
This contract also demonstrates a situation where a nonnegative integer might need to be set to a negative value
(if, for example, a user's collateral value falls suddenly, and the user's borrowing is no longer overcollateralized).

\iffalse

As such, every transaction on the protocol recomputes the maximum remaining amount a user can borrow in that block,
and sets the value of the constraint (via the $set(\cdot)$ method)
to this value, before adding and subtracting to this value (as the user performs actions).
Two transactions in the same block from the same user will compute the same initial constraint,
and therefore this constraint does not cause a conflict.

\fi

\paragraph{API Changes}
The main API change that we make to Compound
regards tracking the protocol's total borrowing and lending.
Compound computes variable interest rates based on the amounts of the base asset
borrowed from and supplied to the contract.  
Our implementation uses contract-wide nonnegative integers
to compute these values, but to update these counters, each transaction
must know whether a transfer of the base asset out of the account is borrowing more or supplying less.
In \SCS{}, a transaction cannot determine this on its own, in the presence of concurrent transfers.

The Ethereum implementation stores a user's base asset balance as a single, signed integer.  
In contrast,
our implementation splits a user's base asset balance into two nonnegative integer counters, one for borrowing and one for supplying.
Transactions then specify whether a transfer is to reduce supply or to borrow more (or some combination).
This could result in a user simultaneously borrowing and supplying, so we also add a method to cancel out this situation.

\subsection{Sequencers and the Actor Model}
\label{sec:sequencers}

The ordered sets of \S \ref{sec:sets} can recreate sequential semantics when required.
In each block, a contract can create a new set, and transactions can
submit an action (i.e., a function call) to this set.  
Then, in a subsequent block, one transaction can sequentially
load and sequentially apply all of the
actions from a previous block.
Some decentralized exchanges, such as Serum \cite{serumtechnical} and 
SundaeSwap \cite{sundaeswapscalability}, use this type of mechanism already, 
even when the underlying blockchain guarantees strictly serializable
semantics.

Our implementation of this mechanism sorts accumlated actions by a fee paid
upon submission of an action,
and pays out this fee to whoever executes accumulated events.

\iffalse
The implementation allocates one ordered
set per block, in which events are accumulated.
%to which $add\_event(\cdot)$ inserts elements.  
Then, at a later block, anyone can iterate over and apply all events accumulatd in the earlier block.%call $exec\_round(\cdot)$ to apply all of the accumulated events.  
Our implementation sorts
accumulated events by a fee bid, which is collected and paid out to
whoever executes the events.
\fi

%whoever calls 
%$exec\_round(\cdot)$.

If every contract were gated by such a sequencer,
 then all contracts would only 
interact via asynchronous message-passing (the Actor Model \cite{hewitt1973session}),
a model for smart contracts deployed in CosmWasm \cite{cosmwasmactor} and 
Near 
\cite{nearcrosscontract}.

In other words, \SCS{}'s semantics are at least as expressive as the
semantics used in production today by two top-50
(by market capitalization \cite{coingecko2022december}) blockchains.  
However, \SCS{}
is strictly more powerful than the message-passing model 
because it enables contracts that do not need 
sequencing to call one another directly---which is especially
important for fundamental, commonly-used contracts such as
token contracts.

\subsubsection{Example: Constant Function Market Makers}
\label{sec:cfmm}

We used the sequencer paradigm to implement a Constant Function Market
Maker \cite{uniswapv2,angeris2020improved}.  Users submit \textit{actions}
(make a trade, deposit liquidity, or withdraw liquidity).  
%All of the
%operations submitted in one \SCS{} block are gathered into the same
%ordered set.  
Users pay fees upon sending actions to the market;
these are collected and paid out to whoever sends the transaction
that applies the actions, as in ~\cite{serumtechnical}.
Actions are applied in order of fee paid.

%Our implementation requires users to pay fees on their
%transactions.  Fees are collected by whoever sends the transaction
%that executes the set of operations, as in~\cite{serumtechnical}.  The
%fee paid is used as the tag on the set (thereby sorting operations by
%fee).

The main complexity that \SCS{} introduces 
is that the runtime cannot reject
individual actions that are applied on the exchange in the same transaction.
If any action on the exchange fails due to a conflict with a concurrent transaction
(e.g., overdrafting an account),
then \SCS{} rejects the entire transaction and does not apply any
actions in the set.

As such, our implementation performs actions that could cause a conflict (such as withdrawing
from an account) 
in the transaction that submits an action, not during the transaction that applies the action.
Our implementation moves assets to the contract to cover a trade when a user submits a trade request,
not when applying the action.
Funds are only disbursed from the contract when actions are applied.

\section{Evaluation: Scalability}
\label{sec:scalability}

We compare the throughput and scalability of \SCS{} against Block-STM
\cite{gelashvili2023block}, an approach (currently deployed in production
on the Aptos blockchain) based on optimistic
concurrency control for deterministic execution of blocks of
transactions on a replicated state machine.  Like \SCS{}, Block-STM
executes transactions that contain untrusted, arbitrary
code and targets financial applications.

% a state of the art
%approach for batched transaction execution based on optimistic concurrency control.  Like \SCS{}, Block-STM focuses on 
%blockchain applications---specifically, Block-STM also deterministically executes batches of transactions.  This optimistic
%approach is deployed in production in the Aptos blockchain \cite{aptosblockstm}.

Our experiments run on one c6a.metal instance in an Amazon Web Services datacenter.
The system has two 48-core AMD EPYC 7R13 processors and 384 GB of memory, with hyperthreading disabled.

\paragraph{Payments Workload}
To control for transaction complexity, the measurements of this section use payment transactions, modeled
on the ``Aptos p2p'' transactions of~\cite{gelashvili2023block}. 
These transactions
read from 8 addresses and write to 5, while \SCS{}'s implementation of this functionality
uses 3 adjustments to nonnegative integers (in the token contract of \S\ref{sec:token}), 
two reads of configuration data,
one set insertion, and one set clear (for the replay prevention
mechanism of \S\ref{sec:replay}). % \ref{fig:replaymechanism}).
This roughly translates to reading 6 addresses and writing to 4.  Each transaction verifies an authorization signature.
In any system,
transactions that require more computation would run more slowly and reduce throughput numbers accordingly.
When serialized, each transaction consumes $196$ bytes.

Replay prevention sets (\S\ref{sec:replay}) are expanded from their default capacity ($64$)
to their maximum ($65535$) if the number of accounts is $10,000$ or smaller.
Our set implementation uses $4$ bytes for every slot of capacity,
and our hardware lacks memory for 10 million maximum-capacity sets.
But to reach a large block size when the number of accounts is small,
each account must send a large number of transactions in a block.

%except when the number of accounts is $10,000$ or smaller,
%when we expand them to their maximum capacity ($65535$).  Our hash set implementation (\S \ref{sec:sets}) 
%uses $4$ bytes for every slot of capacity,
%and our hardware does not have enough memory to support 10 million accounts with maximum capacity sets
%(this would require over 2 terabytes of memory). 
%However, when the number of accounts is small,
%each account must be able to send a large number of transactions in a batch 
%(to reach a batch size of $100,000$).

Each transaction is a payment between two accounts, chosen uniformly at random.
Varying the number of accounts therefore varies the level of contention between transactions.

\paragraph{Varying Contention}

\begin{figure}  
\centering
\includegraphics[width=\columnwidth]{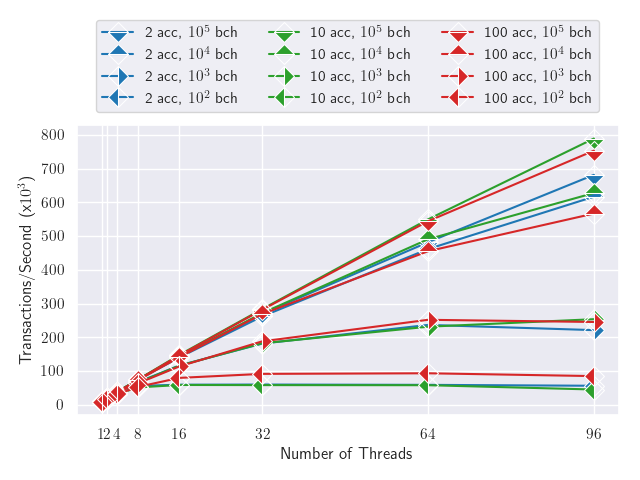}
\caption{
    Transactions per second on \SCS{}, varying the number of accounts ("acc") and batch size ("bch").
    \label{fig:vary_nacc_batch}
}
\end{figure}

Fig. \ref{fig:vary_nacc_batch} plots the end to end transaction throughput rate of \SCS{} proposing (and executing) new blocks of payment transactions,
varying block size and the number of accounts.
The measurements are averages over 20 trials, after 5 warmup rounds. % These experiments vary the number of accounts and the size of each block.
%Replay prevention sets are expanded to their maximum capacity ($65535$).
There is a certain amount of work that must be performed once per block; larger blocks amortize this less effectively parallelizable work over
more transactions.  

%When serialized, each transaction is $196$ bytes, so $100,000$ transactions per second would need a network capable at minimum of
%handling at least $20$ MB/s of sustained traffic.

What makes \SCS{} unique is that the level of contention between transactions does not significantly affect throughput.
When every payment transaction is between the same two accounts, then every transaction modifies the same
addresses.  Under sequential semantics, all
transactions would conflict and
execution would be serialized.  However, \SCS{} achieves the same throughput in this high-contention setting
as in a low contention setting. 

\begin{figure}  
\centering
\includegraphics[width=\columnwidth]{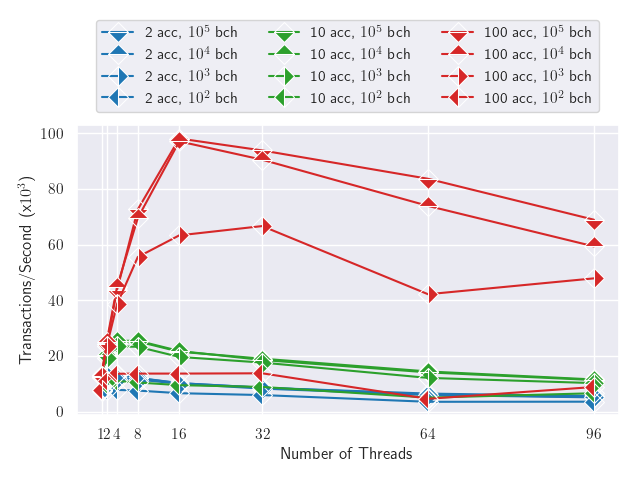}
\caption{
    Transactions per second on Block-STM \cite{gelashvili2023block}, varying the number of accounts ("acc") and batch size ("bch").
    \label{fig:vary_nacc_batch_bstm}
}
\end{figure}

By contrast, Fig.~\ref{fig:vary_nacc_batch_bstm} plots the throughput of Block-STM on the same hardware and transaction patterns.
This design is based on optimistic concurrency control, which must fall back to nearly sequential execution
under high contention (while still paying for concurrency control overhead).

\iffalse

(while still paying the overhead of the optimistic concurrency control --- Figure 7, \cite{gelashvili2023block}).
By contrast, in \SCS{}, all transactions commute by design, so throughput is essentially unaffected by the frequency at which
transactions modify the same locations.  This is why \SCS{} can get linear scalability on workloads that otherwise cannot be scaled.
\TODO{is this ok?} These throughput numbers are comparable to those of Ramseyer et al.~\cite{ramseyer2023speedex}, who take a related approach
 (but only implement a few types of transactions).
 \fi

Two real-world individuals are unlikely to send bursts of thousands of
transactions between each other; however, single accounts often
receive bursts of payment traffic.  Large e-commerce websites or the
Internal Revenue Service, for example, occasionally receive a high
volume of payments, and payroll services may send salary payments to
many employees in a short period of time.  \SCS{}'s commutative
semantics allow institutions to build scalability into their payments
infrastructure and guarantee that their users are not affected by
contented data elsewhere in the system.

\iffalse
add hash to set
clear hashset up to threshold

load registered pk read 1
check signature

load address of token read 1

adjust allowance (write int64) rw 1
adjust balance self (write int64)
adjust balance receiver (write int64)
\fi

\paragraph{Scalability to Many Accounts}

\begin{figure}
\centering
\includegraphics[width=\columnwidth]{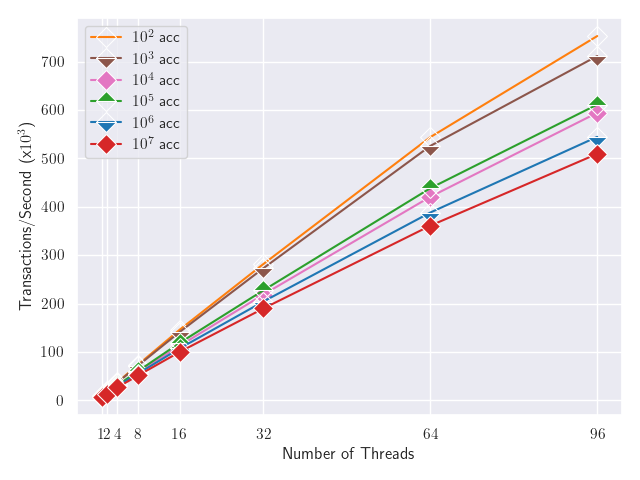}
\caption{
    Transactions per second on \SCS{} with batches of size 100,000, varying the number of accounts.
    \label{fig:vary_nacc_largebatch}
}
\end{figure}

The cost of access to a user's account data increases as the number of users increases.
%As the number of users in a system increases, so too does the cost of each access to a user's account data.
Fig. \ref{fig:vary_nacc_largebatch} 
plots the throughput of \SCS{} as the number of accounts increases, with a fixed batch size of $100,000$.

%Figure \ref{fig:vary_nacc_batch} uses very few active accounts, 
%so as to directly mirror the parameter choices in the measurements of~\cite{gelashvili2022block}.  However,
%a real digital currency system would have far more active users.  Figure \ref{fig:vary_nacc_largebatch} 
%plots the throughput of \SCS{} as the number of active accounts increases, with a fixed batch size of $100,000$.

%The replay prevention sets (\S\ref{sec:replay}) are at their default capacity ($64$), 
%except when the number of accounts is 1,000 or smaller,
%when we expand them to their maximum capacity ($65535$).  Our hash set implementation (\S \ref{sec:sets}) 
%uses $4$ bytes for every slot of capacity,
%and our hardware does not have enough memory to support 10 million accounts with maximum capacity sets
%(this would require over 2 terabytes of memory). 
%However, when the number of accounts is small,
%each account must be able to send a large number of transactions in a batch 
%(to reach a batch size of $100,000$).

\SCS{} throughput decreases as the number of users increases, largely due to the increased lookup time per
account.  The single-threaded throughput with 10 million accounts, for example, is $91\%$ of
the throughput with 10 thousand accounts.  More accounts also slows down the work that is done in each block
after iterating over the transactions, slightly reducing scalability.
The throughput with 96 threads and 10 million accounts is, for
example, $86\%$ of that with 10 thousand accounts (a $73\times$
speedup vs.\ a $78\times$ speedup over the single-threaded benchmark, respectively).

We do not try to optimize \SCS{} for the details of our machine's NUMA architecture; we suspect that some of the scalability decrease
comes from increased variability of memory access times when all cores are active. 

\begin{figure}[t]
\centering
\includegraphics[width=\columnwidth]{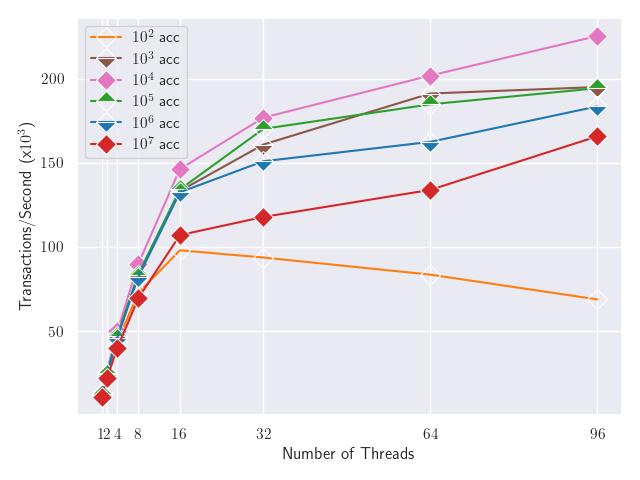}
\caption{
    Transactions per second on Block-STM with batches of size 100,000, varying the number of accounts.
    \label{fig:vary_nacc_largebatch_bstm}
}
\end{figure}

Fig. \ref{fig:vary_nacc_largebatch_bstm} plots the throughput of Block-STM on this workload.
Like \SCS{}, Block-STM's throughput falls as the number of accounts increases.
However, Block-STM's scalability is limited, even in extreme cases where there are millions of accounts
and there is virtually no contention between transactions.  The overhead of concurrency control appears to become the
dominant runtime factor, even when there are no concurrency conflicts.

\begin{figure}[t]
\centering
\includegraphics[width=\columnwidth]{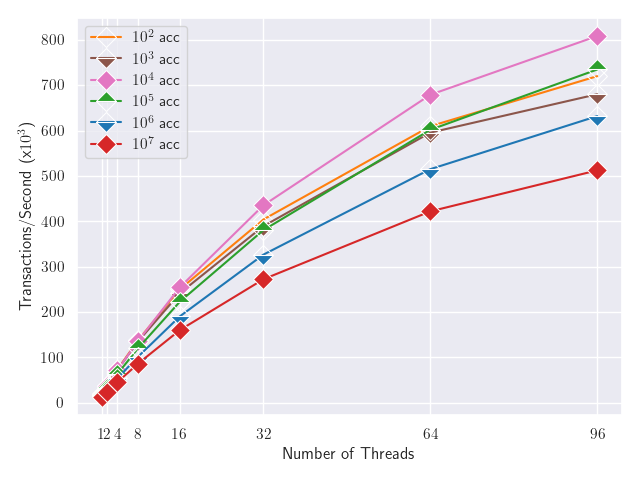}
\caption{
    Payments per second on SPEEDEX \cite{ramseyer2023speedex} with batches of size 100,000, varying the number of accounts.
    \label{fig:vary_nacc_largebatch_speedex}
}
\end{figure}

As another comparison,
SPEEDEX~\cite{ramseyer2023speedex} also uses a design based on parallel execution of semantically concurrent transactions.
However, SPEEDEX only supports a small set of fixed operations, specialized for one application,
whereas transactions in \SCS{} can execute arbitrary, untrusted computation.  Fig. \ref{fig:vary_nacc_largebatch_speedex}
plots the throughput of SPEEDEX on an analogous payment workload.

%Naturally, as the number of active accounts grows larger, so too does the lookup time per each storage access
%(reducing absolute throughput).  The single-threaded throughput with 10 million accounts, for example, is $88\%$ of
%the throughput with 10 thousand accounts.  More active accounts also slows down the work that is done in each batch
%after iterating over the transactions, so more accounts lead to slightly worse scalability results.
%The throughput with 96 threads and 10 million accounts is, for example, $69\%$ of that with 10 thousand accounts (an $80\times$ speedup vs a $64\times$ 
%speedup over the single-threaded benchmark, respectively)

%We do not try to optimize \SCS{} for the precise details of our machine's NUMA architecture; we suspect that some of the scalability decrease
%comes from increased variability of memory access times when all cores are active. 
%The single largest source of throughput loss under higher thread counts
%is the work of building the Merkle trie that logs which storage locations are modified (see \S \ref{sec:implementation}).
% Each thread builds a threadlocal trie as it modifies persistent storage, and the process of merging these tries
%after a batch is not as scalable as the rest of \SCS{}.  Nodes in these tries are allocated from threadlocal buffers, which might also interact 
%poorly with
%the system's NUMA architecture.
%\TODO{the modified keys trie log merge is extremely variable in its runtime, for reasons I don't fully understand, at 96 threads.}

\paragraph{Data Persistence Overhead}
The c6a.metal instances in the previous experiments do not have persistent data storage devices
and accordingly the experiments (on both \SCS{} and Block-STM) do not log data to persistent storage.
%We instrumented our \SCS{} implementation to log all state changes to disk (\S \ref{sec:data_structures})
We enabled logging in our \SCS{} implementation to record state
changes persistently (\S \ref{sec:data_structures})
and ran the experiments again on a c6id.metal Amazon Web Services instances,
which had two 32-core Intel Xeon Platinum 8375C processors with hyperthreading disabled,
256 GB of memory,
and 4x1900GB NVMe drives in a RAID 0 configuration.

Unsurprisingly, logging overhead increases as the number of open accounts increases
and as the number of worker threads increases.
Peak overhead was $3\%$, $6\%$, and $9\%$, in the parameter settings with $64$ threads
and $10^5$, $10^6$, and $10^7$ accounts, respectively.  Without logging,
throughput reached $476k$, $424k$, and $383k$ transactions per second, respectively.
We push the writes to disk off of the critical path;
the overhead comes from preparing the data to write to disk. 

\paragraph{Conclusions}
More important than any topline number is \SCS{}'s scalability.  
We implement (\S \ref{sec:impl_short}) what we believe to be a minimal complete set of features.
However, any deployed system may want extra features, each of which adds overhead.
%For example, transactions might require multiple authorization signatures, or an implementation might choose
%a slower but simpler (and easier to audit) implementation of the WebAssembly sandbox.  

Sequential semantics force a deployment to choose between more features
and higher throughput.
\SCS{} adds a dimension of cost to this tradeoff.
A deployment can compensate for additional features by adding additional compute hardware.
It may be unlikely that any near to medium term deployment of a digital currency would need multiple hundreds of thousands
of transactions per second, but this scalability enables a deployment to smoothly compensate for future feature demands and increased workload.

These measurements only count the rate of successful transactions.  
If many transactions sent to \SCS{} are invalid,
a block proposer would execute these transactions but not include them in a proposal.  
The measured throughput would fall accordingly,
but the scalability property of \SCS{} would be unaffected.
% (in fact, the rejected transactions do not influence
%the per-block overhead).
%To put these numbers in context, it may be quite unlikely that any near to medium term deployment of a digital currency would need multiple hundreds of thousands
%of transactions per second.  

\iffalse
The key factor that we demonstrate here is that \SCS{} is scalable.  More important than any top-line number is the fact 
that more compute resources let \SCS{} process more transactions per second.

As discussed in \S \ref{sec:implementation}, we focus here on implementing the features of a deployed system that 
require specific interaction with \SCS{}.  
Every feature that one would add to a deployed system would likely add a small amount of overhead to each transaction.
More complicated authorization checks on each transaction (such as requiring multiple signatures) slow down each transaction.
A sequentially operating state machine gives a deployment an inherent tradeoff between more features (and slower transactions,
meaning lower throughput) and faster transactions (and fewer features).  \SCS{}'s adds a dimension of cost to this tradeoff.
A deployment can compensate for the throughput loss that results from adding a feature by simply adding additional compute hardware.

\fi

\section{Extension: Incomplete External Calls}
\label{sec:rpc}

\SCS{}'s semantics enable behavior not implementable in a sequentially
executing state machine.  When assembling a new block proposal, \SCS{}
can pause a transaction and reallocate resources elsewhere, without
disrupting the execution of other transactions.  This enables
transactions in \SCS{} to query external services for signed (or
otherwise self-authenticating) results.  If an external call fails to
terminate within a time bound, \SCS{} can drop the transaction without
affecting other transactions or wasting additional CPU time.
Replicability requires the block proposer to include the results of
external calls in its block proposal.

%Replication requires that the results of external calls be
%self-authenticating (e.g., signed by a relevant authority).  
%(so this feature is not compatible with a leaderless consensus protocol).
Access to external information is an important business in existing
blockchains, but suffers from high latency, as
queries can only return results in subsequent blocks.  These calls in
\SCS{} let contracts directly access external information; for
example, a bank might hold personal information privately and
authorize transfers based on queries to this data without revealing
the private information.

\paragraph{Atomic Swaps}
External calls enable efficient off-chain market-making, 
 addressing the so-called ``free option'' problem of existing decentralized exchanges that execute
 atomic swaps~\cite{han2019optionality,poon2017plasma}.  
 In an atomic swap, two users agree to a trade between two
assets at some exchange rate.  Implementing this requires making two payments in one transaction,
and requires authorization from both users.  Typically, one user signs
the transaction first, then sends it to the other,
who signs it and submits it to the blockchain.

However, upon receipt of the first user's signature, the second user has the option of dropping the transaction.
To minimize swap failures due to network delay or blockchain congestion, swap transactions often include a relatively long expiration time.
The optionality benefits the second user at the expense of the first~\cite{han2019optionality}.

By contrast, in \SCS{}, if a swap is initiated within a transaction
via a call to an external service (i.e., a request for quote API),
both swap participants know that if the swap settles, then it will settle within the same block---within the smallest discrete amount of time
that a blockchain can measure.  The \SCS{} replica has the option to cancel the swap, but neither of the swap participants do.

\section{Discussion and Limitations}
\label{sec:limitations}

\paragraph{Snapshot Reads}

\SCS{} provides a different notion of consistency than is typically used in distributed databases.
These semantics may be unusual, but they mirror the way users interact with
blockchains today.  Most users, when deciding whether to send a transaction, 
are only able to see the state of the blockchain
at the close of the most recent block.  Any preflight analysis, 
therefore, executes against this snapshot. %, without seeing any modifications made by other transactions in the block.
Differences between a user's snapshot and the state that a transaction reads in a sequential system
have caused real-world security challenges \cite{erc20attack}.

%The potential for differences between a user's snapshot view of state and sequential semantics
%has caused security challenges in real-world systems \cite{erc20attack}.
%\TODO{get a citation on this}
%In practice, many users of existing systems
%pay significant fees to ensure their transactions are included at the start of a (sequentially executed) block,
%so as to ensure that transactions execute as expected (e.g., to capture an arbitrage opportunity). 

These semantics allow non-serializable execution traces, such as write-skew anomalies.
If necessary, \SCS{} places the responsibility for
implementing serializability onto the contract.  Contracts can implement their own locks (\S \ref{sec:locks})
on their internal state.
%For example, these semantics admit write-skew anomalies, where one transaction might read two variables $X$ and $Y$, 
%and then write to $Y$,
%while another transaction in the same block reads the same two variables and writes to $X$.
%If such behavior is incompatible with the contract's higher-level application requirements,
%then the transactions should acquire a lock on $X$ and $Y$ (so only one could be included in a block).

Unlike designs based on eventual consistency and conflict-free replicated data
types (CRDTs)~\cite{shapiro2011conflict}, \SCS{}'s data structures need not support arbitrarily delayed updates, 
%from transactions arbitrarily far in the past, 
but rather,
only concurrent updates from the same block.
This property is what allows \SCS{} to implement account balances and the nonnegative integer object (\S \ref{sec:nnint}).
Furthermore, state machines can synchronize with each other at block
boundaries (as in, e.g., \S \ref{sec:sequencers}),
 without \SCS{} explicitly tracking dependency chains.
%This property is crucial to implement, for example, asset transfers between account balances.
By contrast, CRDTs cannot maintain an 
integer counter that enforces a nonnegativity constraint and allows subtraction.
%However, because a transaction in block $k$ is guaranteed to see all of the outputs of transactions that were in block $k-1$ and none from transactions
%in block $k+1$, \SCS{}'s runtime can implement this primitive (\S \ref{sec:nnint}) by looking only at a finite set of transactions.

\paragraph{Development Difficulty}

The execution semantics of \SCS{} may make developing applications more difficult than in a
sequential system.
Anecdotally, we found it helpful to view a contract's API as a set of actions that a caller might perform,
subject to a set of constraints. 
We conjecture that not all but a vast majority of important applications
can be built in a manner that admits some concurrency.

As discussed in \S\ref{sec:sequencers},
production blockchains today~\cite{cosmwasmactor,nearcrosscontract} use semantics, based on asynchronous message passing,
that are strictly weaker than those of \SCS{}.  
This gives a baseline implementation 
for any contract used in those systems.  \SCS{}'s block assembly process
introduces some potential pitfalls in this message-passing model,
which we discuss in \S \ref{sec:cfmm}.

Key applications that \SCS{} fundamentally cannot implement other than in this baseline actor model
are exchanges and automated market-makers (\S \ref{sec:cfmm}).
Ramseyer et al.~\cite{ramseyer2023speedex} show a different way of operating an exchange
in a design like that of \SCS{} based on blocks of concurrent transactions.
However, \cite{ramseyer2023speedex} relies an application-specific optimization framework
to clear blocks of trades, which a smart contract in \SCS{} cannot implement at scale.

\paragraph{Leader-based Design}
\SCS{}'s semantics eliminate the ability of one replica to manipulate transaction ordering within a block, 
but a malicious block proposer might delay transactions.
%Censorship and delay resistance are properties of a consensus protocol, not the execution engine.
We envision
\SCS{} in an environment with large financial institutions operating replicas, where regulation
ensures non-malicious operation in most cases.  \SCS{}'s 
leader-based reserve-commit process may be incompatible some leaderless consensus protocols~\cite{danezis2022narwhal,keidar2021all}
that are more censorship-resistant.

%The challenge is in maintaining object constraints
%The challenge is that \SCS{} relies on its reserve-commit process to maintain
%constraints.
%The challenge is that \SCS{} cannot maintain object constraints if transactions
%in a batch conflict. 
If \SCS{} is given as input a new block (from the consensus protocol,
instead of using the $\reserve()$-$\commit()$ process) 
that causes a constraint violation,
it would need to prune out sufficient transactions to prevent violations.
We do not implement it, but \SCS{} could use a generalization of the deterministic 
filtering mechanism in ~(\cite{ramseyer2023speedex}, Apx. I).

Specifically, \SCS{} could execute all transactions in a block,
and identify any objects on which transactions conflict or that are put in a constraint-violating state.
\SCS{} can then compute, for each object, a ``conflict set,''
such that removing an object's conflict set ensures that the object is not put in a constraint-violating state.
For example, if concurrent transactions write different values to a bytestring, then all of these transactions are in the conflict set.
If a nonnegative integer's constraint is broken, then all transactions that subtract from its value form the conflict set.
Removing transactions cannot cause additional conflicts in \SCS{},
so removing the union of all conflict sets
leaves a block with no constraint violations.

\paragraph{Lock Prioritization}

\SCS{} does not implement functionality to allow some transactions to have higher-priority access to locks (\S \ref{sec:locks}).
Some smart contracts could therefore be vulnerable to denial of service attacks, through an excess of transactions
acquiring internal locks.  \SCS{} places the responsibility for imposing costs on this behavior to the smart contract developer.
As a simple example, a smart contract could require that any
transaction that acquires a lock also perform an action.

Importantly, \SCS{} does not begin to call $\reserve()$ methods until the WebAssembly module has terminated execution in a successful state.
A contract can rely on this guarantee to force transactions that acquire locks to somehow pay for them.
This prevents attacks where, for example, a transaction acquires a
lock, performs an expensive computation, and then aborts.  (Such transactions
are no-ops, and are dropped prior to line~\ref{line:start_rc_loop} of
Algorithm~\ref{alg:parallel_exec}.)

\section{Related Work}

\paragraph{Semantic Changes}
Our approach is modeled on the Scalable Commutativity Rule of Clements
et al.~\cite{clements2015scalable}, who build scalability into an operating system
by designing commutative system call semantics.  Ramseyer et al.~\cite{ramseyer2023speedex} design commutative semantics
specific to a decentralized exchange; our work generalizes their approach to arbitrary smart contracts.
Walther~\cite{walther2019optimization}, CowSwap ~\cite{cowswapproblem}, and Penumbra~\cite{penumbraswap} 
also design specific applications
that execute transactions in unordered batches.

Conflict-free Replicated Data Types \cite{shapiro2011conflict} give commutative interfaces to data structures in a distributed environment,
but support arbitrarily delayed operations.
They cannot, therefore, implement a nonnegative account balance that admits subtraction.
 Garamv{\"o}lgyi et al.~\cite{garamvolgyi2022utilizing}, resolve some write conflicts
with a ``commutative add'' instruction.  However, this design maintains sequential semantics,
and does not support commutative subtraction from a counter.
Studies of historical Ethereum data ~\cite{saraph2019empirical,garamvolgyi2022utilizing} find that most conflicts
come from integer counters. 

%As such, they cannot implement, for example, allow subtraction from an account balance
%while maintaining a nonnegativity constraint.

\paragraph{Parallel Execution}

Block-STM \cite{gelashvili2023block} optimistically executes transactions in parallel, re-executing in case of conflict.
This meets our requirements for a digital currency engine, except that its throughput does not scale
under limited contention.
Chen et al.~\cite{chen2021forerunner} speculatively execute Ethereum transactions, and Lin et al.~\cite{lin2022operation} instrument Ethereum execution traces to reduce re-execution overhead after concurrency conflicts.

Eve \cite{kapritsos2012all} executes batches of transactions concurrently, but relies on an application-specific conflict checker to 
assemble batches of nonconflicting transactions.  This checker requires advance knowledge of transaction read and write sets,
and performance degrades substantially when this checker produces errors.
Malicious users could attack any conflict checker with malicious code to introduce errors.
By contrast, transactions in \SCS{} execute arbitrary, untrusted code, and \SCS{} needs no advance knowledge of read or write sets.
%A more subtle challenge for this approach in a digital currency setting is that it is possible, albeit unlikely,
%that every replica would execute the 

LiTM \cite{xia2019litm} takes a batch of transactions and executes on a single snapshot as many nonconflicting transactions as possible,
then recomputes a new state snapshot and continues.  However, throughput is limited by round frequency,
because a data item 
can only be updated once per round,
and Gelashvili et al.~\cite{gelashvili2023block}
found throughput scalability of LiTM to be limited under contention.

Strife \cite{prasaad2020handling} dynamically splits a batch of transactions
into partitions where transactions in different partitions do not conflict, and executes these subsets in parallel.
Bohm \cite{faleiro2014rethinking} deterministically executes batches of transactions using a multi-version data structure. 
Both approaches
rely on advance knowledge of transaction read and write sets.

Basil \cite{suri2021basil} builds a key-value store in a context where replicas and users may be faulty,
boosting performance by adapting optimistic techniques to this setting.  However, the system relies on users
to execute transactions, which means that it cannot ensure correct execution or prevent
Byzantine clients from overwriting the data of correct users.

%in their case at around 170k transactions per second.
% higher throughput requried
%a 2-phase commit style system like \SCS{}.

%citations here not as important, can be removed
Other approaches
compute nonconflicting transaction sets~\cite{bartoletti2020true,yu2018parallel,stellartxfootprintcap} or
include conflict resolution information in block proposals~\cite{dickerson2019adding,anjana2019efficient,anjana2020efficient}. 
Ruan et al.~\cite{ruan2020transactional} reorder transactions to reduce conflict rates.
%and Hyperledger Fabric
%invalidates conflicting transaction pairs \cite{hyperledgerfabric}.

\paragraph{Ordering Semantics}

Li et al.~\cite{li2012making,li2014automating} propose RedBlue consistency, in which a transaction 
can be broken into noncommutative (Red) and commutative (Blue)
components.  Red components must still be executed sequentially, limiting scalability, and Blue components may or may not be visible to 
Red ones, making system operation nondeterministic.  PoR consistency \cite{li2018fine} generalizes this notion to more fine-grained
ordering requirements, but has the same challenges in our context.
Systems like COPS~\cite{lloyd2011don} likewise relax consistency guarantees
to increase performance but do not guarantee reproducible operation.

\SCS{} deliberately places no
ordering between transactions in a block.  Other decentralized systems
build a ``fair ordering'' assuming a bounded fraction of malicious
replicas~\cite{kelkar2020order,zhang2020byzantine,ramseyer2023fair}
or commit to an ordering before
revealing transaction contents~\cite{clineclockwork,zhang2022flash,schmid2021secure}.

%allow transactions to be tagged
%as commutative in a distributed database application.  Many other relaxations of sequentially consistent semantics have been implemented
%~\cite{lloyd2011don,li2018fine,herlihy1991wait}

%with 
%implement specific state machines with commutative semantics in a blockchain context.

\paragraph{Divided State}
Several blockchain scaling approaches divide the system into
independently-running state machines (``shards'')~\cite{skychain,polkadot,nearsharding,das2020efficient} 
or side-chains (``Layer-2
networks'')~\cite{poon2016bitcoin,kalodner2018arbitrum,poon2017plasma,loopringdesigndoc}.
%polygon,goel2020continuous,zkrollup,optimisticrollup}.
Each component runs independently, but between components, these designs complicate interaction and limit
throughput.  BIDL \cite{qi2021bidl} provides a total order on transactions among a (permissioned) group of organizations,
and each organization only executes transactions relevant to it (which complicates cross-organization transactions).  However, each organization executes transactions sequentially, on one thread,
and the system is not programmable as in \SCS{}---each transaction executes one of a set of predefined functions, implemented as precompiled code
(thereby avoiding the overhead of sandboxing untrusted user code).

%BIDL executes transactions sequentially, and each ``organization'' only executes the subset of transactions relevant to it.
%This mean that transactions cannot operate atomically across organization boundaries (although of course organizations can use the common leader/sequencing
%of BIDL to pass messages, and implement e.g. an actor-based model or 2phase-commit).
%BIDL also leaves transaction semantics up to each organization -- as far as we can infer,
%BIDL nodes execute transactions with precompiled machine code (regular go code, 
%https://github.com/hku-systems/bidl/blob/main/bidl/normal_node/cmd/server/core/execute.go#L126 and #L213).  
%By contrast,
%Hedgehog allows untrusted clients to publish code, and for security it must sandbox the code (in our case, with a WASM interpreter).
%This sandboxing adds significant overhead to each transaction, and makes scalability all the more important.

Some designs operate each smart contract independently (in parallel), allowing communication between contracts only through asynchronous
messages \cite{cosmwasmactor,nearcrosscontract,kaleem2021event}.  
%scilla2018 is not asynchronous -- the contracts are implemented in an actor model,
% but everything executes as one atomic unit (sequentially)
% Or so it seems -- the 2019 oopsla paper says sequential, the whitepaper makes weird claims about sharding
This model complicates many common tasks, such as asset transfers (which would ideally be done atomically), as each call to a token contract is asynchronous.

Several systems, such as Bitcoin~\cite{nakamoto:bitcoin} and Project Hamilton~\cite{hamilton}, divide user balances into pieces, where each piece is an ``Unspent Transaction Output'' (UTXO).  These systems enable some parallelism, as in ~\cite{bagaria2019prism},
but complicate the user experience and abstractions built on UTXOs ~\cite{sundaeswapscalability}.
%Use of UTXOs enables some parallelism (as in
%e.g. \cite{bagaria2019prism}), but complicates the user experience---each user must track
%a set of UTXOs, not  one account balance---and complicates abstractions built on them ~\cite{sundaeswap}
Project Hamilton~\cite{hamilton} furthermore found that computing a total ordering on transactions became a throughput bottleneck.
\SCS{}, by contrast, scales without this tradeoff.

\section{Conclusion}

We presented \SCS{}, our novel design for a smart contract execution
engine.  \SCS{} is designed to execute blocks of transactions in
parallel, but unlike prior work, \SCS{} semantically places no
ordering between transactions in the same block.

\SCS{}'s key insight is to design a set of commutative semantics
that allow the \SCS{} runtime to deterministically resolve concurrent writes
to the same object.  To support contexts where transactions may irresolvably conflict
(such as when withdrawing money from an account), \SCS{} lets contracts express constraints
about their data objects.  \SCS{} maintains these constraints with a reserve-commit process
to assemble blocks that contain no conflicts.

These semantics enable \SCS{} to execute transactions in parallel with minimal synchronization overhead
between threads.  This parallelism allows \SCS{} to scale its transaction throughput as it is given more compute resources,
a property which would be crucial for a real-world digital currency system.
Furthermore, the commutativity ensures that the rate at which transactions modify the same objects (which would
lead to conflicts in designs based on optimistic concurrency control) does not meaningfully affect throughput.

Our implementation of \SCS{} achieves more than half a million payments per second between 10M accounts, but
achieves even more than that when transacting between only 2 acccounts.

%-------------------------------------------------------------------------------
\bibliographystyle{plain}
\bibliography{main}

\begin{thebibliography}{10}

\bibitem{cowswapproblem}
Cow protocol overview: The batch auction optimization problem.
\newblock
  \url{https://web.archive.org/web/20220614183101/https://docs.cow.fi/off-chain-services/in-depth-solver-specification/the-batch-auction-optimization-problem}.
\newblock Accessed 10/19/2022.

\bibitem{dai}
The maker protocol: Makerdao’s multi-collateral dai (mcd) system.
\newblock \url{https://makerdao.com/en/whitepaper/}.
\newblock Accessed 12/14/2021.

\bibitem{penumbraswap}
The penumbra protocol: Sealed-bid batch swaps.
\newblock
  \url{https://web.archive.org/web/20220614034906/https://protocol.penumbra.zone/main/zswap/swap.html}.
\newblock Accessed 10/19/2022.

\bibitem{solanarent}
Solana documentation: Rent.
\newblock
  \url{https://web.archive.org/web/20220903121658/https://docs.solana.com/implemented-proposals/rent}.

\bibitem{serumtechnical}
A technical introduction to the serum dex.
\newblock \url{
  https://web.archive.org/web/20221112173940/https://docs.google.com/document/d/1isGJES4jzQutI0GtQGuqtrBUqeHxl_xJNXdtOv4SdII/edit
  }.
\newblock Accessed 12/11/22.

\bibitem{wasm3}
wasm3: The fastest webassembly interpreter, and the most universal runtime.
\newblock \url { https://github.com/wasm3/wasm3 }.

\bibitem{loopringdesigndoc}
Loopring 3 design doc.
\newblock
  \url{https://web.archive.org/web/20220411224154/https://github.com/Loopring/protocols/blob/master/packages/loopring_v3/DESIGN.md#results},
  2021.

\bibitem{compoundv3}
Compound iii docs.
\newblock \url{https://docs.compound.finance/}, 2022.

\bibitem{cosmwasmactor}
Cosmwasm documentation: Actor model for contract calls.
\newblock
  \url{https://web.archive.org/web/20220811195538/https://docs.cosmwasm.com/docs/1.0/architecture/actor/},
  2022.

\bibitem{nearcrosscontract}
Near documentation: Cross-contract calls.
\newblock
  \url{https://web.archive.org/web/20221021051657/https://docs.near.org/develop/contracts/crosscontract},
  2022.

\bibitem{wasminstrument}
Wasm-instrument: Instrument and transform wasm modules.
\newblock \url{https://github.com/paritytech/wasm-instrument}, 2022.

\bibitem{compoundapp}
Compound finance.
\newblock
  \url{https://web.archive.org/web/20230414212550/https://compound.finance/},
  2023.
\newblock Accessed 04/14/2023.

\bibitem{uniswapv2}
Hayden Adams, Noah Zinsmeister, and Dan Robinson.
\newblock Uniswap v2 core.
\newblock 2020.

\bibitem{angeris2020improved}
Guillermo Angeris and Tarun Chitra.
\newblock Improved price oracles: Constant function market makers.
\newblock In {\em Proceedings of the 2nd ACM Conference on Advances in
  Financial Technologies}, pages 80--91, 2020.

\bibitem{anjana2020efficient}
Parwat~Singh Anjana, Hagit Attiya, Sweta Kumari, Sathya Peri, and Archit
  Somani.
\newblock Efficient concurrent execution of smart contracts in blockchains
  using object-based transactional memory.
\newblock In {\em International Conference on Networked Systems}, pages 77--93.
  Springer, 2020.

\bibitem{anjana2019efficient}
Parwat~Singh Anjana, Sweta Kumari, Sathya Peri, Sachin Rathor, and Archit
  Somani.
\newblock An efficient framework for optimistic concurrent execution of smart
  contracts.
\newblock In {\em 2019 27th Euromicro International Conference on Parallel,
  Distributed and Network-Based Processing (PDP)}, pages 83--92. IEEE, 2019.

\bibitem{bagaria2019prism}
Vivek Bagaria, Sreeram Kannan, David Tse, Giulia Fanti, and Pramod Viswanath.
\newblock Prism: Deconstructing the blockchain to approach physical limits.
\newblock In {\em Proceedings of the 2019 ACM SIGSAC Conference on Computer and
  Communications Security}, pages 585--602, 2019.

\bibitem{bartoletti2020true}
Massimo Bartoletti, Letterio Galletta, and Maurizio Murgia.
\newblock A true concurrent model of smart contracts executions.
\newblock In {\em International Conference on Coordination Languages and
  Models}, pages 243--260. Springer, 2020.

\bibitem{ethstateexpiry}
Vitalik Buterin.
\newblock A state expiry and statelessness roadmap.
\newblock
  \url{https://web.archive.org/web/20220916204724/https://notes.ethereum.org/@vbuterin/verkle_and_state_expiry_proposal},
  2021.

\bibitem{castro:bfs}
Miguel Castro and Barbara Liskov.
\newblock Practical byzantine fault tolerance.
\newblock In {\em 3rd Symposium on Operating Systems Design and
  Implementation}, pages 173--186, New Orleans, LA, February 1999.

\bibitem{chen2021forerunner}
Yang Chen, Zhongxin Guo, Runhuai Li, Shuo Chen, Lidong Zhou, Yajin Zhou, and
  Xian Zhang.
\newblock Forerunner: Constraint-based speculative transaction execution for
  ethereum (full version).
\newblock 2021.

\bibitem{clements2015scalable}
Austin~T Clements, M~Frans Kaashoek, Nickolai Zeldovich, Robert~T Morris, and
  Eddie Kohler.
\newblock The scalable commutativity rule: Designing scalable software for
  multicore processors.
\newblock {\em ACM Transactions on Computer Systems (TOCS)}, 32(4):1--47, 2015.

\bibitem{clineclockwork}
Dan Cline, Thaddeus Dryja, and Neha Narula.
\newblock Clockwork: An exchange protocol for proofs of non front-running.

\bibitem{coingecko2022december}
CoinGecko.
\newblock Cryptocurrency prices by market cap.
\newblock
  \url{https://web.archive.org/web/20221210015628/https://www.coingecko.com/}.

\bibitem{daian2019flash}
Philip Daian, Steven Goldfeder, Tyler Kell, Yunqi Li, Xueyuan Zhao, Iddo
  Bentov, Lorenz Breidenbach, and Ari Juels.
\newblock Flash boys 2.0: Frontrunning, transaction reordering, and consensus
  instability in decentralized exchanges.
\newblock {\em arXiv preprint arXiv:1904.05234}, 2019.

\bibitem{danezis2022narwhal}
George Danezis, Lefteris Kokoris-Kogias, Alberto Sonnino, and Alexander
  Spiegelman.
\newblock Narwhal and tusk: a dag-based mempool and efficient bft consensus.
\newblock In {\em Proceedings of the Seventeenth European Conference on
  Computer Systems}, pages 34--50, 2022.

\bibitem{das2020efficient}
Sourav Das, Vinith Krishnan, and Ling Ren.
\newblock Efficient cross-shard transaction execution in sharded blockchains.
\newblock {\em arXiv preprint arXiv:2007.14521}, 2020.

\bibitem{dechev2010understanding}
Damian Dechev, Peter Pirkelbauer, and Bjarne Stroustrup.
\newblock Understanding and effectively preventing the aba problem in
  descriptor-based lock-free designs.
\newblock In {\em 2010 13th IEEE international symposium on
  object/component/service-oriented real-time distributed computing}, pages
  185--192. IEEE, 2010.

\bibitem{dickerson2019adding}
Thomas Dickerson, Paul Gazzillo, Maurice Herlihy, and Eric Koskinen.
\newblock Adding concurrency to smart contracts.
\newblock {\em Distributed Computing}, pages 1--17, 2019.

\bibitem{erc20openzeppelin}
OpenZeppelin Documentation.
\newblock Erc20.
\newblock \url{https://docs.openzeppelin.com/contracts/2.x/api/token/erc20}.
\newblock Accessed 12/10/22.

\bibitem{faleiro2014rethinking}
Jose~M Faleiro and Daniel~J Abadi.
\newblock Rethinking serializable multiversion concurrency control.
\newblock {\em arXiv preprint arXiv:1412.2324}, 2014.

\bibitem{garamvolgyi2022utilizing}
P{\'e}ter Garamv{\"o}lgyi, Yuxi Liu, Dong Zhou, Fan Long, and Ming Wu.
\newblock Utilizing parallelism in smart contracts on decentralized blockchains
  by taming application-inherent conflicts.
\newblock {\em arXiv preprint arXiv:2201.03749}, 2022.

\bibitem{gelashvili2023block}
Rati Gelashvili, Alexander Spiegelman, Zhuolun Xiang, George Danezis, Zekun Li,
  Dahlia Malkhi, Yu~Xia, and Runtian Zhou.
\newblock Block-stm: Scaling blockchain execution by turning ordering curse to
  a performance blessing.
\newblock In {\em Proceedings of the 28th ACM SIGPLAN Annual Symposium on
  Principles and Practice of Parallel Programming}, pages 232--244, 2023.

\bibitem{gilad:algorand}
Yossi Gilad, Rotem Hemo, Silvio Micali, Georgios Vlachos, and Nickolai
  Zeldovich.
\newblock Algorand: Scaling byzantine agreements for cryptocurrencies.
\newblock In {\em Proceedings of the 26th Symposium on Operating Systems
  Principles}, SOSP '17, page 51–68, New York, NY, USA, 2017. Association for
  Computing Machinery.

\bibitem{han2019optionality}
Runchao Han, Haoyu Lin, and Jiangshan Yu.
\newblock On the optionality and fairness of atomic swaps.
\newblock In {\em Proceedings of the 1st ACM Conference on Advances in
  Financial Technologies}, pages 62--75, 2019.

\bibitem{hewitt1973session}
Carl Hewitt, Peter Bishop, and Richard Steiger.
\newblock A universal modular actor formalism for artificial intelligence.
\newblock In {\em Advance Papers of the Conference}, volume~3, page 235.
  Stanford Research Institute Menlo Park, CA, 1973.

\bibitem{stellartxfootprintcap}
Graydon Hoare.
\newblock Core advancement protocol 46-05: Smart contract data.
\newblock
  \url{https://web.archive.org/web/20221212214501/https://github.com/stellar/stellar-protocol/blob/master/core/cap-0046-05.md},
  May 2022.

\bibitem{kaleem2021event}
Mudabbir Kaleem, Keshav Kasichainula, Rabimba Karanjai, Lei Xu, Zhimin Gao, Lin
  Chen, and Weidong Shi.
\newblock An event driven framework for smart contract execution.
\newblock In {\em Proceedings of the 15th ACM International Conference on
  Distributed and Event-based Systems}, pages 78--89, 2021.

\bibitem{kalodner2018arbitrum}
Harry Kalodner, Steven Goldfeder, Xiaoqi Chen, S~Matthew Weinberg, and Edward~W
  Felten.
\newblock Arbitrum: Scalable, private smart contracts.
\newblock In {\em 27th $\{$USENIX$\}$ Security Symposium ($\{$USENIX$\}$
  Security 18)}, pages 1353--1370, 2018.

\bibitem{kapritsos2012all}
Manos Kapritsos, Yang Wang, Vivien Quema, Allen Clement, Lorenzo Alvisi, and
  Mike Dahlin.
\newblock All about eve: Execute-verify replication for multi-core servers.
\newblock In {\em Presented as part of the 10th $\{$USENIX$\}$ Symposium on
  Operating Systems Design and Implementation ($\{$OSDI$\}$ 12)}, pages
  237--250, 2012.

\bibitem{keidar2021all}
Idit Keidar, Eleftherios Kokoris-Kogias, Oded Naor, and Alexander Spiegelman.
\newblock All you need is dag.
\newblock In {\em Proceedings of the 2021 ACM Symposium on Principles of
  Distributed Computing}, pages 165--175, 2021.

\bibitem{kelkar2020order}
Mahimna Kelkar, Fan Zhang, Steven Goldfeder, and Ari Juels.
\newblock Order-fairness for byzantine consensus.
\newblock In {\em Annual International Cryptology Conference}, pages 451--480.
  Springer, 2020.

\bibitem{leshner2019compound}
Robert Leshner and Geoffrey Hayes.
\newblock Compound: The money market protocol.
\newblock {\em White Paper}, 2019.

\bibitem{li2014automating}
Cheng Li, Jo{\~a}o Leit{\~a}o, Allen Clement, Nuno Pregui{\c{c}}a, Rodrigo
  Rodrigues, and Viktor Vafeiadis.
\newblock Automating the choice of consistency levels in replicated systems.
\newblock In {\em 2014 USENIX Annual Technical Conference (USENIX ATC 14)},
  pages 281--292, 2014.

\bibitem{li2012making}
Cheng Li, Daniel Porto, Allen Clement, Johannes Gehrke, Nuno Pregui{\c{c}}a,
  and Rodrigo Rodrigues.
\newblock Making $\{$Geo-Replicated$\}$ systems fast as possible, consistent
  when necessary.
\newblock In {\em 10th USENIX Symposium on Operating Systems Design and
  Implementation (OSDI 12)}, pages 265--278, 2012.

\bibitem{li2018fine}
Cheng Li, Nuno Pregui{\c{c}}a, and Rodrigo Rodrigues.
\newblock Fine-grained consistency for geo-replicated systems.
\newblock In {\em 2018 USENIX Annual Technical Conference (USENIX ATC 18)},
  pages 359--372, 2018.

\bibitem{lin2022operation}
Haoran Lin, Yajin Zhou, and Lei Wu.
\newblock Operation-level concurrent transaction execution for blockchains.
\newblock {\em arXiv preprint arXiv:2211.07911}, 2022.

\bibitem{lloyd2011don}
Wyatt Lloyd, Michael~J Freedman, Michael Kaminsky, and David~G Andersen.
\newblock Don't settle for eventual: Scalable causal consistency for wide-area
  storage with cops.
\newblock In {\em Proceedings of the Twenty-Third ACM Symposium on Operating
  Systems Principles}, pages 401--416, 2011.

\bibitem{lokhava:stellar}
Marta Lokhava, Giuliano Losa, David Mazi\`{e}res, Graydon Hoare, Nicolas Barry,
  Eli Gafni, Jonathan Jove, Rafa\l{} Malinowsky, and Jed McCaleb.
\newblock Fast and secure global payments with stellar.
\newblock In {\em Proceedings of the 27th ACM Symposium on Operating Systems
  Principles}, SOSP '19, page 80–96, New York, NY, USA, 2019. Association for
  Computing Machinery.

\bibitem{hamilton}
James Lovejoy, Madars Virza, Cory Fields, Kevin Karwaski, Anders Brownworth,
  and Neha Narula.
\newblock Hamilton: A {High-Performance} transaction processor for central bank
  digital currencies.
\newblock In {\em 20th USENIX Symposium on Networked Systems Design and
  Implementation (NSDI 23)}, pages 901--915, Boston, MA, April 2023. USENIX
  Association.

\bibitem{nakamoto:bitcoin}
Satoshi Nakamoto.
\newblock Bitcoin: A peer-to-peer electronic cash system, 2008.
\newblock \url{http://bitcoin.org/bitcoin.pdf}.

\bibitem{poon2017plasma}
Joseph Poon and Vitalik Buterin.
\newblock Plasma: Scalable autonomous smart contracts.
\newblock {\em White paper}, pages 1--47, 2017.

\bibitem{poon2016bitcoin}
Joseph Poon and Thaddeus Dryja.
\newblock The bitcoin lightning network: Scalable off-chain instant payments,
  2016.

\bibitem{prasaad2020handling}
Guna Prasaad, Alvin Cheung, and Dan Suciu.
\newblock Handling highly contended oltp workloads using fast dynamic
  partitioning.
\newblock In {\em Proceedings of the 2020 ACM SIGMOD International Conference
  on Management of Data}, pages 527--542, 2020.

\bibitem{qi2021bidl}
Ji~Qi, Xusheng Chen, Yunpeng Jiang, Jianyu Jiang, Tianxiang Shen, Shixiong
  Zhao, Sen Wang, Gong Zhang, Li~Chen, Man~Ho Au, and Heming Cui.
\newblock Bidl: A high-throughput, low-latency permissioned blockchain
  framework for datacenter networks.
\newblock In {\em Proceedings of the ACM SIGOPS 28th Symposium on Operating
  Systems Principles}, SOSP '21, page 18–34, New York, NY, USA, 2021.
  Association for Computing Machinery.

\bibitem{qin2022quantifying}
Kaihua Qin, Liyi Zhou, and Arthur Gervais.
\newblock Quantifying blockchain extractable value: How dark is the forest?
\newblock In {\em 2022 IEEE Symposium on Security and Privacy (SP)}, pages
  198--214. IEEE, 2022.

\bibitem{ramseyer2023fair}
Geoffrey Ramseyer and Ashish Goel.
\newblock Fair ordering via streaming social choice theory.
\newblock {\em arXiv preprint arXiv:2304.02730}, 2023.

\bibitem{ramseyer2023speedex}
Geoffrey Ramseyer, Ashish Goel, and David Mazi{\`e}res.
\newblock {SPEEDEX}: A scalable, parallelizable, and economically efficient
  decentralized {EXchange}.
\newblock In {\em 20th USENIX Symposium on Networked Systems Design and
  Implementation (NSDI 23)}, pages 849--875, Boston, MA, April 2023. USENIX
  Association.

\bibitem{ruan2020transactional}
Pingcheng Ruan, Dumitrel Loghin, Quang-Trung Ta, Meihui Zhang, Gang Chen, and
  Beng~Chin Ooi.
\newblock A transactional perspective on execute-order-validate blockchains.
\newblock In {\em Proceedings of the 2020 ACM SIGMOD International Conference
  on Management of Data}, pages 543--557, 2020.

\bibitem{saraph2019empirical}
Vikram Saraph and Maurice Herlihy.
\newblock An empirical study of speculative concurrency in ethereum smart
  contracts.
\newblock {\em arXiv preprint arXiv:1901.01376}, 2019.

\bibitem{solanaoutage}
Leopold Schabel.
\newblock Reflections on solana's sept 14 outage.
\newblock
  \url{https://web.archive.org/web/20211104012332/https://jumpcrypto.com/reflections-on-the-sept-14-solana-outage/},
  Oct 2021.
\newblock Accessed 12/7/2021.

\bibitem{schmid2021secure}
Noah Schmid, Christian Cachin, Orestis Alpos, and Giorgia Marson.
\newblock Secure causal atomic broadcast, 2021.

\bibitem{shapiro2011conflict}
Marc Shapiro, Nuno Pregui{\c{c}}a, Carlos Baquero, and Marek Zawirski.
\newblock Conflict-free replicated data types.
\newblock In {\em Symposium on Self-Stabilizing Systems}, pages 386--400.
  Springer, 2011.

\bibitem{suri2021basil}
Florian Suri-Payer, Matthew Burke, Zheng Wang, Yunhao Zhang, Lorenzo Alvisi,
  and Natacha Crooks.
\newblock Basil: Breaking up bft with acid (transactions).
\newblock In {\em Proceedings of the ACM SIGOPS 28th Symposium on Operating
  Systems Principles}, pages 1--17, 2021.

\bibitem{nearsharding}
NEAR Team.
\newblock Near launches nightshade sharding, paving the way for mass adoption.
\newblock
  \url{https://web.archive.org/web/20221007081239/https://near.org/blog/near-launches-nightshade-sharding-paving-the-way-for-mass-adoption/},
  November 2021.
\newblock Accessed 10/18/2022.

\bibitem{sundaeswapscalability}
SundaeSwap Team.
\newblock Sundaeswap scalability.
\newblock
  \url{https://web.archive.org/web/20220523234648/https://sundaeswap.finance/posts/sundaeswap-scalability},
  November 2021.

\bibitem{erc20attack}
Mikhail Vladimirov and Dmitry Khovratovich.
\newblock Erc20 api: An attack vector on the approve/transferfrom methods.
\newblock
  \url{https://web.archive.org/web/20221108114451/https://docs.google.com/document/d/1YLPtQxZu1UAvO9cZ1O2RPXBbT0mooh4DYKjA_jp-RLM/edit}.

\bibitem{eip20}
Fabian Vogelsteller and Vitalik Buterin.
\newblock Eip 20: Erc-20 token standard.
\newblock {\em Ethereum Improvement Proposals}, 20, 2015.

\bibitem{walther2019optimization}
Tom Walther.
\newblock An optimization model for multi-asset batch auctions with uniform
  clearing prices.
\newblock In {\em Operations Research Proceedings 2018: Selected Papers of the
  Annual International Conference of the German Operations Research Society
  (GOR), Brussels, Belgium, September 12-14, 2018}, pages 225--231. Springer,
  2019.

\bibitem{warren20170x}
Will Warren and Amir Bandeali.
\newblock 0x: An open protocol for decentralized exchange on the ethereum
  blockchain.
\newblock 2017.

\bibitem{polkadot}
Gavin Wood.
\newblock Polkadot: Vision for a heterogeneous multi-chain framework.
\newblock {\em White Paper}, 21, 2016.

\bibitem{xia2019litm}
Yu~Xia, Xiangyao Yu, William Moses, Julian Shun, and Srinivas Devadas.
\newblock Litm: a lightweight deterministic software transactional memory
  system.
\newblock In {\em Proceedings of the 10th International Workshop on Programming
  Models and Applications for Multicores and Manycores}, pages 1--10, 2019.

\bibitem{yin:hotstuff}
Maofan Yin, Dahlia Malkhi, Michael~K. Reiter, Guy~Golan Gueta, and Ittai
  Abraham.
\newblock Hotstuff: Bft consensus with linearity and responsiveness.
\newblock In {\em Proceedings of the 2019 ACM Symposium on Principles of
  Distributed Computing}, PODC '19, page 347–356, New York, NY, USA, 2019.
  Association for Computing Machinery.

\bibitem{yu2018parallel}
Wei Yu, Kan Luo, Yi~Ding, Guang You, and Kai Hu.
\newblock A parallel smart contract model.
\newblock In {\em Proceedings of the 2018 International Conference on Machine
  Learning and Machine Intelligence}, pages 72--77, 2018.

\bibitem{zhang2022flash}
Haoqian Zhang, Louis-Henri Merino, Vero Estrada-Galinanes, and Bryan Ford.
\newblock Flash freezing flash boys: Countering blockchain front-running.
\newblock In {\em The Workshop on Decentralized Internet, Networks, Protocols,
  and Systems (DINPS)}, 2022.

\bibitem{skychain}
Jianting Zhang, Zicong Hong, Xiaoyu Qiu, Yufeng Zhan, Song Guo, and Wuhui Chen.
\newblock Skychain: A deep reinforcement learning-empowered dynamic blockchain
  sharding system.
\newblock In {\em 49th International Conference on Parallel Processing-ICPP},
  pages 1--11, 2020.

\bibitem{zhang2020byzantine}
Yunhao Zhang, Srinath Setty, Qi~Chen, Lidong Zhou, and Lorenzo Alvisi.
\newblock Byzantine ordered consensus without byzantine oligarchy.
\newblock In {\em 14th $\{$USENIX$\}$ Symposium on Operating Systems Design and
  Implementation ($\{$OSDI$\}$ 20)}, pages 633--649, 2020.

\end{thebibliography}

%%%%%%%%%%%%%%%%%%%%%%%%%%%%%%%%%%%%%%%%%%%%%%%%%%%%%%%%%%%%%%%%%%%%%%%%%%%%%%%%
\end{document}